\documentclass[sigplan,10pt,screen,nonacm]{acmart}
\usepackage{graphicx}
\usepackage{subcaption}
\usepackage{dsfont}
\usepackage{braket}
\usepackage{tikz}
\usepackage{listings}
\usepackage{mathtools}
\usepackage{dblfloatfix}

\newcommand{\code}[1]{\texttt{#1}}

\makeatletter
\lstdefinelanguage{llvm}{
  morecomment = [l]{;},
  morecomment = [l]{//},
  morestring=[b]",
  sensitive = true,
  classoffset=0,
  keywordstyle=\ttfamily\color{black},
  morekeywords={},
  classoffset=1,
  keywordstyle=\ttfamily\color[rgb]{0.7,0.7,0.9},
  morekeywords={
    to, step
  },
  classoffset=2, keywordstyle=\bfseries\color[rgb]{0,0,0.6},
  morekeywords={
    for, if, else, and, addi, subi, muli, index_cast, shift_left, shift_right_signed, shift_right_unsigned, cmpi, constant, func, call, load, store, uitofp, log2, ceilf, fptoui, divf, return, cond_br, divi_unsigned, br, yield, H, X, R, CX, SWAP, alloc, allocreg, extract, combine, circ, getval, adj, ctrl, apply, free, freereg, meas, phi
  },
  alsoletter={\%},
  keywordsprefix={\%},
}
\makeatother

\begin{document}

\title{Enabling Dataflow Optimization for Quantum Programs}

\author{David Ittah}
\affiliation{\institution{ETH Zurich}}
\email{david.ittah@mail.mcgill.ca}

\author{Thomas Häner}
\affiliation{\institution{Microsoft Quantum}}
\email{thhaner@microsoft.com}

\author{Vadym Kliuchnikov}
\affiliation{\institution{Microsoft Quantum}}
\email{v.kliuchnikov@gmail.com}

\author{Torsten Hoefler}
\affiliation{\institution{ETH Zurich}}
\email{htor@inf.ethz.ch}

\begin{abstract}
We propose an IR for quantum computing that directly exposes quantum and classical data dependencies for the purpose of optimization. The \emph{Quantum Intermediate Representation for Optimization} (QIRO) consists of two dialects, one input dialect and one that is specifically tailored to enable quantum-classical co-optimization. While the first employs a perhaps more intuitive memory-semantics (quantum operations act as side-effects), the latter uses value-semantics (operations consume and produce states). Crucially, this encodes the dataflow directly in the IR, allowing for a host of optimizations that leverage dataflow analysis. We discuss how to map existing quantum programming languages to the input dialect and how to lower the resulting IR to the optimization dialect. We present a prototype implementation based on MLIR that includes several quantum-specific optimization passes. Our benchmarks show that significant improvements in resource requirements are possible even through static optimization. In contrast to circuit optimization at run time, this is achieved while incurring only a small constant overhead in compilation time, making this a compelling approach for quantum program optimization at application scale.
\end{abstract}


\maketitle


\section{Introduction}
In recent years, the quantum programming landscape has seen a boom of new languages, tools, and environments \cite{qiskit19, projectq18, quil16, qsharp18, silq2020, quipper13, scaffold12, scaffcc14, cirq2020, strawberryfields19, qwire17}. The focus of these projects varies widely, ranging from low-level interfaces for prototype hardware \cite{qiskit19, quil16} to high-level algorithm development \cite{quipper13, qsharp18, silq2020}. When tasked with supporting common components of the compilation stack, these projects have opted for one of two approaches; either (1) a complete (re-)implementation of these components or (2) re-use of existing infrastructure by embedding the domain-specific language in a widely-used programming language such as Python. Indeed, most quantum programming languages that target Noisy Intermediate-Scale Quantum (NISQ)~\cite{preskill18} hardware are embedded in a classical programming language. Such embedded domain-specific languages (eDSLs) can be viewed as libraries that generate a data structure representing a quantum circuit. These data structures can be seen as very simple and flat intermediate representations (IRs) without any control flow.

In light of the large number of quantum gates required to achieve practical quantum speedups, however, it is clear that a more advanced IR is needed to support large-scale algorithms. Indeed, many chemistry applications of practical interest require between $10^9$ and $10^{15}$ gates \cite{reiher16,vonburg2020} and applications in the domain of cryptography similarly lead to programs with $10^{10}$ gates \cite{haner2020elliptic,gidney19}. Consequently using flat data structures such as lists of gates\footnote{Representing a quantum program as a flat list of gates is similar to representing a classical program as a list of ALU instructions without any control flow.} or directed acyclic graphs (DAGs) to represent quantum programs is infeasible at application scale. Instead, a special purpose IR that incorporates classical and quantum control flow is needed.

Designing a more advanced IR for quantum programs poses some unique challenges not present in classical computing. Most notably, ``values'' held in qubits cannot be copied due to the no-cloning theorem of quantum mechanics~\cite[Box 12.1]{nielsen02}. Existing quantum IRs such as the recently proposed QIR~\cite{qir2020} therefore opt to only represent \textit{references} to quantum data, while operations on qubits are modeled via side-effects. Unfortunately, such a representation severely limits reuse of existing compiler components, which often rely on the dataflow to be explicit in the IR. As a remedy, we introduce a quantum-analog of SSA\footnote{Static Single Assignment (SSA) is an IR property where each variable is assigned exactly once, see Section~\ref{sec:ir-background}.} where dataflow is explicit.

\begin{figure*}[t]
\includegraphics[width=\linewidth]{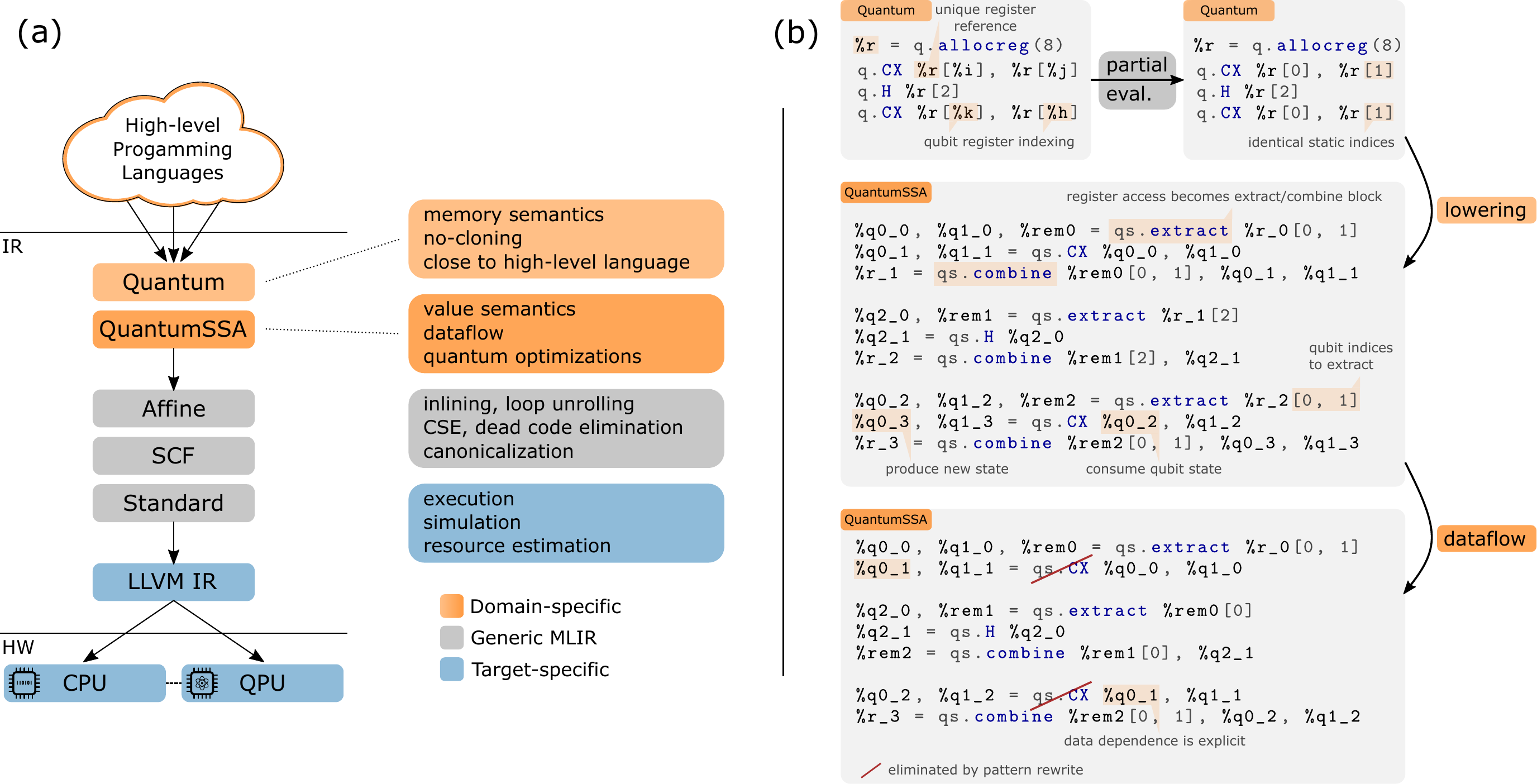}
\caption{(a) Proposed compilation stack incorporating our IR infrastructure, along with the features and main concepts of the different levels.
(b) In the input dialect (Quantum), operations act on unique qubit references via side-effects (memory-semantics). We designed our optimization dialect to facilitate dataflow analysis and to maximally leverage existing optimization infrastructure. Therefore, operations in the optimization dialect (QuantumSSA) consume and return quantum states (value-semantics) and quantum register accesses are translated to pairs of extract/combine instructions.}
\label{fig:main}
\end{figure*}

The no-cloning theorem might also seem to rule out a large set of classical program optimizations that rely on the duplication of values across the program such as common sub-expression elimination (CSE). However, such techniques are useful for optimization of mixed quantum-classical programs, as illustrated by the example in Figure~\ref{fig:main}b: While two successive CNOT (commonly CX) gates may be eliminated from the IR if they are applied to the same qubits, it is difficult to carry out such an optimization in the presence of dynamic accesses to quantum registers, especially if dataflow is not explicit in the IR. However, if existing optimization passes such as partial evaluation succeed at inferring, e.g., that \code{i=k=0} and \code{j=h=1}, then the two CNOTs may be removed by leveraging our optimization dialect, which directly exposes (quantum and classical) dataflow. Classical program optimization thus immediately increases the usefulness of quantum-specific optimization passes.

\subsection{Quantum Multi-Level IR}
In this work, we introduce the \emph{Quantum Intermediate Representation for Optimization} (QIRO), an IR for universal quantum computation that leverages MLIR~\cite{mlir2021} to support quantum-classical co-optimi\-zation. In contrast to existing IRs for quantum computing, we design our optimization dialect in a way such that data dependencies are explicit for both quantum and classical variables. This enables the use of existing infrastructure as well as the development of future quantum-specific optimization passes that leverage dataflow analysis. Figure~\ref{fig:main}b depicts how a code fragment is translated from our input dialect (\emph{Quantum}) to the optimization dialect (\emph{QuantumSSA}). An IR-specific optimization transforms the code in the middle section to that of the bottom section by consolidating \texttt{combine} and \texttt{extract} instructions, which are used to represent quantum register accesses in the optimization dialect. Optimizing such patterns in the IR enables subsequent optimizations (such as the CX-CX elimination) by exposing quantum data dependencies inside registers.

The use of MLIR presents significant benefits over a single-level IR such as LLVM. For example, our approach leverages several application-specific abstractions from MLIR such as classical (low-level) computation, domain-specific quantum computation, and specialized affine loop representations. Furthermore, language or domain-specific optimizations are comparatively difficult to implement in LLVM due to a lack of high-level information. LLVM's rigidness implies that domain-specific operations must be represented by opaque functions, and additional types by opaque pointers (the approach taken by QIR, see Section~\ref{sec:related}). In contrast, MLIR provides multiple abstractions via which to interact with the IR on a high level, even across application domains. This includes, for instance, extensible types that can be mixed and matched, operation attributes for compile-time information, and operation traits for the reuse of passes in an extensible system (see Section \ref{sec:background}).

Finally, MLIR features an extensive validation mechanism for operations, types, and traits. We leverage this mechanism to express invariants of the IR and to statically enforce constraints on quantum operations wherever possible.

\smallskip
\subsection{Contributions}
We design and implement QIRO, a novel IR for quantum-classical co-optimization in MLIR. Our dual dialect approach guarantees simple translation from high-level quantum programming languages as well as broad re-use of existing compilation infrastructure for optimization.

In short, our main contributions are:
\begin{itemize}
    \item We propose an optimization dialect, which can be seen as a quantum-analog of SSA. The optimization dialect (1) is compatible with the no-cloning theorem and (2) allows for reuse of existing compiler components (quantum and classical dataflow is explicit in the IR).
    \item We define a higher-level IR that serves as an input dialect with the same semantics that are commonly used for quantum IRs, so-called \textit{memory-semantics} (quantum operations are modeled as side-effects).
    \item We leverage existing work (MLIR) to implement the two proposed dialects, including the lowering from the input dialect to the optimization dialect.
    \item We describe how a variety of existing quantum programming languages can be mapped to our input dialect.
    \item We show that our IR enables optimization and resource estimation at application-scale up to 5-6 orders of magnitude faster than existing frameworks that rely on optimization at run time such as Qiskit~\cite{qiskit19} and ProjectQ~\cite{projectq18}.
    \item We implement several optimization passes that aim to reduce the quantum resource requirements (operations and qubits). We demonstrate that for Shor's algorithm, practically all ($\sim$99.8\%) savings identified by ProjectQ's run-time circuit optimization may be obtained statically, at significantly lower cost in terms of optimizer run time.
\end{itemize}

By directly exposing quantum and classical dataflow, QIRO enables future quantum and quantum-classical optimizations that make use of this information. Combined with its advantages in terms of compilation time, our IR may serve as a helpful tool for resource estimation of optimized quantum-classical programs. In turn, this allows for more efficient hardware-software co-design and for achieving a quantum advantage for real-world problems once fault-tolerant quantum computers become available.

\section{Background}
\label{sec:background}
This section provides a short introduction to classical IRs, compiler optimization, and quantum computing. For more in-depth treatments of these subjects, we refer the reader to the textbooks by \citet{aho86} and \citet{nielsen02}, respectively.

\subsection{Intermediate Representation}\label{sec:ir-background}
In general, intermediate representations (IR) are extremely useful for the implementation of multi-language / multi-architecture compiler suites, as well as simplifying and enhancing verification, analysis, and optimization tasks. While various forms of intermediate representations exist, SSA and SSA-based IRs have played a key role in the design of our approach, so we will briefly introduce these concepts below.

\paragraph{Static Single Assignment}
A common property of intermediate representations in compilers of imperative languages, SSA form mandates that every value is assigned exactly once (in a static sense), while imposing no restrictions on the number of uses. In order to handle assignments to a variable from different control paths, IRs traditionally rely on the use of a pseudo-operation called the $\phi$-function. At points of control flow merges, these $\phi$-functions represent the selection of the correct value from a set of assignments according to the actual execution path. This encodes the \emph{static} uncertainty of where a value in use was defined, while also obeying the single-definition rule. The code listing below illustrates the use of $\phi$-functions with a translation from pseudo-code (left) to SSA-form (right).
\smallskip
\begin{center}
\begin{minipage}{3cm}
\begin{lstlisting}[language=llvm,frame=tb,numbers=left,xleftmargin=0.5cm]
if (cond)
    x = 6
else
    x = 4
y = x*2
\end{lstlisting}
\end{minipage}
$~\longrightarrow~$
\begin{minipage}{3.5cm}
\begin{lstlisting}[language=llvm,frame=tb,numbers=left,xleftmargin=0.5cm]
if (cond)
    x1 = 6
else
    x2 = 4
x3 = phi(x1,x2)
y1 = x3*2
\end{lstlisting}
\end{minipage}
\end{center}
 
 Having a representation in SSA form is of great advantage to certain optimizations as data dependencies are explicit in the program structure. Reaching definitions analysis\footnote{A definition of a variable \textbf{d} reaches a point in the code \textbf{p} if there is an execution path from \textbf{d} to \textbf{p} with no occurrence along that path that ends the validity of d~\cite{aho86}. This form of dataflow analysis becomes obsolete in SSA form since every variable has exactly one definition.} becomes obsolete, def-use graphs remain compact, and dataflow analysis is simple and sparse, all of which are beneficial to optimizations that rely on such information.
 
\paragraph{SSA-based IRs} SSA form has been used for many internal compiler representations, including GCC's GIMPLE, LLVM IR, and others. We briefly describe LLVM as it has had a strong influence on the design of MLIR, which we use in this work.

LLVM \cite{llvm04} models an architecture close to traditional processors with a RISC-like instruction set. It features an infinite set of registers represented by \emph{SSA values} (of primitive type: Boolean, integer, floating-point, pointer), which exposes the dataflow of the program. This allows LLVM to perform transformations without expensive dataflow analysis. Control flow is also explicit in the IR by organizing function bodies into basic blocks, i.e., sections of linear code that always end in a terminator operation that transfers control to another block. Merging control flow with regards to SSA value definitions is handled using explicit $\phi$ instructions, in direct correspondence with theoretical $\phi$-functions.

\subsection{MLIR}
In contrast to the rigidity of LLVM, MLIR~\cite{mlir2021} provides an extensible representation with infrastructure for transformations, analysis, and debugging. There exists a variety of mechanisms to enable and manage extensibility in MLIR.

An extension to MLIR is structured into a \emph{dialect}, akin to a namespace, in which all specialized types, operations, and other IR objects are defined. An \emph{operation} is the fundamental unit of execution in MLIR, similar to instructions in LLVM. Each operation defines its own semantics, allowing dialects to represent constructs at arbitrary levels of abstraction. It is important to note that this freely extensible system need not result in a proliferation of disjoint and self-contained dialects. Instead, dialect components such as operations, but also types and transformations, can be freely mixed and reused across dialects.

Moreover, MLIR provides an extensive IR validation mechanism, with which to verify requirements upon IR construction, invariants across transformations, and so on. Verifiers can be defined at the level of operations, types, and traits. These constructs are vital to the extensible system, and encourage designers to focus on reusable and modular components, without inhibiting dialect-specific implementations where appropriate.

The following table summarizes the syntax of relevant MLIR components described in this section. In order to avoid naming collisions between dialects, operation and type names besides the built-in ones are prefixed with a dialect shorthand.

\begin{center}
\begin{tabular}{l l l}
    \textbf{Construct} & \textbf{Syntax} & \textbf{Example} \\
    \hline
    SSA Value          & \code{\%}         & \code{\%0} \\
    Symbol (e.g. function) & \code{@} & \code{@mod} \\
    Block name         & \code{\^}         & \code{\^{}while} \\
    Dialect Prefix     & \code{dp.}        & \code{std.} \\
    Dialect Operation  & \code{dp.name}    & \code{scf.for} \\
    Dialect Type       & \code{!dp.name}   & \code{!linalg.range}
\end{tabular}
\end{center}
\vspace{\baselineskip}

The MLIR code of a modulo function in Figure~\ref{fig:modulo} will be used to illustrate the concepts that follow. At its core, MLIR employs a functional form of SSA, which distinguishes itself from traditional $\phi$-based SSA forms (such as the one found in LLVM) via the use of \emph{block arguments}. Basic blocks (e.g. L5, L11) that form the nodes in the control-flow graph use block arguments for values that are defined in multiple parent blocks. Every block must end in a \emph{terminator} operation that determines where to transfer control next (e.g. jump, conditional branch (L9), return (L12)). Terminators that transfer control to a block with arguments must provide the desired values, similar to a function call. SSA values in MLIR can appear as either block arguments (e.g. \lstinline!%a_0! on L5), operation operands (e.g. \lstinline{

\begin{figure*}
\begin{lstlisting}[language=llvm,xleftmargin=1cm,xrightmargin=1cm,frame=tb,numbers=left]
func @mod(%a: i64, %N: i64) -> i64 {
    %cond_0 = cmpi "uge", %a, %N : i64
    cond_br %cond_0, ^while(%a: i64), ^ret(%a: i64)

    ^while(%a_0: i64):
        %a_1 = subi %a_0, %N : i64

        %cond_1 = cmpi "uge", %a_1, %N : i64
        cond_br %cond_1, ^while(%a_1: i64), ^ret(%a_1: i64)

    ^ret(%res: i64):
        return %res : i64
}
\end{lstlisting}
\caption{Modulo function implemented in MLIR.}
\label{fig:modulo}
\end{figure*}

Another advantage of MLIR is its capability to represent hierarchical code. Operations can define \emph{regions}, which themselves contain other operations, allowing for arbitrary nesting. A function in MLIR (e.g. \lstinline{func @mod} on L1) is such an operation with a single nested region (L2-L12) containing the function body. As a consequence, loop nests for example need not be represented as linearized control flow via blocks and branches, but can use nested loop operations if more appropriate. The affine dialect makes extensive use of this, and we will later see how this benefits resource estimation (see Section \ref{sec:res-est}). The code listing below illustrates the difference in representation between linearized control flow (top) and structured control flow (bottom), omitting some boilerplate code in computing the loop condition to highlight the structural difference.
\smallskip
\begin{center}
\begin{minipage}{8.3cm}
\begin{lstlisting}[language=llvm,frame=tb,numbers=left,xleftmargin=0.5cm]
%i = constant 0 : index
br ^header
^header:
    cond_br %loop_cond, ^body, ^exit
^body:
    cond_br %if_cond, ^true, ^false
^true:
    ...
    br ^more_body
^false:
    ...
    br ^more_body
^more_body:
    br ^header
^exit:
\end{lstlisting}
\end{minipage}
$~\Updownarrow~$
\begin{minipage}{8.3cm}
\begin{lstlisting}[language=llvm,frame=tb,numbers=left,xleftmargin=0.5cm,aboveskip=0.3cm]
scf.for %i = 0 to 100 {
    scf.if %if_cond {
        ...
    } else {
        ...
    }
}
\end{lstlisting}
\end{minipage}
\end{center}

Several transformation mechanisms are available in MLIR. Every operation can implement specialized hooks for canonicalization and folding purposes. A DAG-to-DAG pattern rewriter simplifies the implementation of transformations that can be expressed as a simple replacement of one DAG pattern with another. A DAG pattern constitutes of a set of operations connected by the usage of values (via arguments) and their definitions (via return values), so-called def-use chains (e.g. the operations \lstinline{subi} (L6) and \lstinline{cmpi} (L8) in Figure~\ref{fig:modulo} are linked via the def-use chain of the value \lstinline!%a_1!). The pattern rewriter can then identify patterns by traversing def-use chains in the IR, and transform any matches accordingly. Besides the pattern rewriting framework, general operation passes can be written that arbitrarily modify an operation and all those nested within. Finally, further infrastructure is provided to simplify dialect lowering and type conversions.

\subsection{Quantum Computing}
\paragraph{Quantum State}
The quantum analog of a classical bit is a quantum bit or qubit. Whereas a classical bit can be in one of two states at any given time, a qubit is in a complex superposition of two basis states $\ket0, \ket1$ corresponding to the values 0 and 1, respectively. The state $\ket\psi$ of a qubit can thus be written as
\[
    \ket\psi = \alpha_0\ket0 + \alpha_1\ket 1,
\]
where $\alpha_0,\alpha_1\in\mathds C$ are complex numbers, so-called \textit{probability amplitudes}, such that $|\alpha_0|^2+|\alpha_1|^2=1$. As the name suggests, these complex numbers are related to probabilities. Namely, measuring a qubit yields a classical bit equal to 0 or 1 with probability $p=|\alpha_0|^2$ or $|\alpha_1|^2=1-p$, respectively. Measurement also collapses the state onto the observed outcome, meaning that the post-measurement state will be $\ket0$ or $\ket1$.

The quantum state $\ket\phi$ of $n$ qubits may be written as a superposition over all $2^n$ $n$-bit strings,
\[
    \ket\phi=\alpha_0\ket{\underbrace{0\cdots0}_n} + \cdots + \alpha_{2^n-1}\ket{\underbrace{1\cdots 1}_n},
\]
where the amplitudes again satisfy the normalization condition $\sum_i|\alpha_i|^2=1$.
Usually, the $n$-bit strings are interpreted as integers, resulting in a shorter notation:
\[
    \ket\phi = \sum_i\alpha_i\ket i.
\]
Measuring all $n$ qubits collapses the state onto $\ket i$ with probability $p_i=|\alpha_i|^2$ and yields the outcome $i$.

In contrast to the state of a classical bit, general quantum states cannot be copied due to the no-cloning theorem~\cite[Box 12.1]{nielsen02}. Specifically, the theorem implies that there exists no unitary operator (see below) $U$ such that $U\ket{\phi}\ket{0}=\ket{\phi}\ket{\phi}$ for all states $\ket{\phi}$.

\paragraph{Quantum Operations}
Similarly to classical computers, the state of a quantum computer may be altered by applying quantum operations to qubits. These operations can be represented as unitary\footnote{Recall that $U$ is unitary if $U^\dagger U=UU^\dagger=\mathds 1$} matrices $U\in\mathds C^{2^n\times 2^n}$, and the state after applying a quantum operation with matrix-representation $U$ is
\[
    \ket{\phi'} = U\ket\phi,
\]
where $\ket\phi$ is interpreted as a column vector of amplitudes $(\alpha_0,\cdots,\alpha_{2^n-1})^T$ and then multiplied with the matrix $U$. The inverse of an operation is defined by its Hermitian adjoint (conjugate transpose), denoted $U^\dagger$. Operations that are hermitian\footnote{Recall that $U$ is hermitian if $U = U^\dagger$} form their own inverses, and commonly appear in optimization techniques.

A few common single-qubit gates are shown in Figure~\ref{fig:gates}, such as the Hadamard gate $H$, the Pauli $X$, $Y$, $Z$ gates, and the $T$ and $S$ gates. Moreover, we use the standard definition from~\citet[Exercise 4.1]{nielsen02} for the rotation gates $R$, $Rx$, $Ry$, $Rz$. Many important multi-qubit gates are controlled versions of single-qubit gates, which apply the single-qubit gate only on the subspace where all control qubits are equal to $\ket1$. An $n$-ary controlled single-qubit gate $U$ can be written as
\[
    ^cU = (\mathds 1-\ket{1\cdots 1}\bra{1\cdots 1})\otimes \mathds 1 + \ket{1\cdots 1}\bra{1\cdots 1}\otimes U,
\]
where $\otimes$ is the Kronecker product and $\ket i\bra i$ denotes the projector onto $\ket i$. For example, the controlled NOT operation (or $CNOT$/$CX$) is a controlled $X$ gate and is also shown in Figure~\ref{fig:gates}.

A set of quantum logic gates $\mathcal{G}$ is termed \emph{universal} if for any unitary $U \in 2^n \times 2^n$ and precision parameter $\epsilon$, there exists a finite sequence of gates $S=G_m...G_2G_1$ where $G_i \in \mathcal{G}$ such that $\max_{\ket{\psi}} ||(S-U)\ket{\psi}|| \leq \epsilon$. That is, $\mathcal{G}$ can be used to approximate $U$ to arbitrary precision. An example of a commonly used universal gate set is the Clifford+T set $\{CNOT, H, S, T\}$.

\begin{figure}
    {\renewcommand{\arraystretch}{2.5}%
    \begin{tabular}{l l l}
    $X = \begin{psmallmatrix}0 & 1\\1 & 0\end{psmallmatrix}$ & $Y = \begin{psmallmatrix}0 & -i\\i & 0\end{psmallmatrix}$ & $Z = \begin{psmallmatrix}1 & 0\\0 & -1\end{psmallmatrix}$\\
    $H = \frac{1}{\sqrt{2}}\begin{psmallmatrix}1 & 1\\1 & -1\end{psmallmatrix}$ & $S = \begin{psmallmatrix}1 & 0\\0 & e^{i\pi/2}\end{psmallmatrix}$ & $T = \begin{psmallmatrix}1 & 0\\0 & e^{i\pi/4}\end{psmallmatrix}$\\
    $R(\theta) = \begin{psmallmatrix}1 & 0\\0 & e^{i\theta}\end{psmallmatrix}$ & \multicolumn{2}{l}{$Rz(\theta)=e^{-i\theta Z /2} = \begin{psmallmatrix}e^{-i\theta/2} & 0\\0 & e^{i\theta/2}\end{psmallmatrix}$}\\
    \multicolumn{3}{l}{$CX = (\mathds 1 - \ket 1\bra 1)\otimes\mathds 1 + \ket 1 \bra 1\otimes X =\begin{psmallmatrix}1&0&0&0\\0&1&0&0\\0&0&0&1\\0&0&1&0\end{psmallmatrix}$} 
    \end{tabular}}
    \caption{Common single- and multi-qubit gates. $Rx$ and $Ry$ (not shown) are defined analogously to $Rz$, with $Rx(\theta)=e^{-i\theta X /2}$ and $Ry(\theta)=e^{-i\theta Y /2}$. The collection of gates shown is a superset of the universal gate set $\{CX, H, S, T\}$.}
    \label{fig:gates}
\end{figure}

\paragraph{Execution Model}
We model quantum computation using a classical ``host'' computer in combination with a quantum co-processor, or quantum processing unit (QPU), with bidirectional real-time communication available. The QPU must be able to support (at minimum) state preparation, a universal gate set, and measurement. A quantum program then consists of a combination of classical and quantum instructions. In each step of the program, the classical host may send sequences of quantum instructions to the quantum co-processor for execution. The responsibility of managing and communicating with the quantum co-processor falls on the quantum runtime environment (RTE), in particular for such tasks as qubit allocation. Consequently, the intricacies of quantum memory management are not considered for the compilation process presented in this work. Sequences of quantum operations may be seen as quantum circuits similar to classical logic circuits. Once the circuit has been executed, the co-processor can return measurement outcomes to the classical host, which may also use these outcomes to alter the execution path.

We distinguish two phases of a program's lifecycle for the purposes of optimization, defined below:
\begin{itemize}
    \item \textbf{compile time:} The program is analyzed and transformed in a fully static way, without the execution of either classical or quantum program parts. In particular, program inputs are unavailable at this stage.
    \item \textbf{run time:} Strictly speaking, this term might be used to describe the execution stage of the program on quantum hardware. However, in the context of optimizations, we also consider any circuit generation phases to be \emph{at run time}. The reason is that such phases already perform ``run-time'' operations such as classical program execution, control flow resolution, and program input propagation, clearly distinguishing it from the static scenario above. The execution of classical meta-programs that perform circuit generation in eDSLs falls into this category.
\end{itemize}

\section{A Quantum Programming Stack}
An overview of the proposed quantum programming stack is presented in Figure~\ref{fig:main}a. We envision that high-level quantum programming languages are translated to QIRO in the proposed input dialect (labeled \textit{Quantum} in the diagram), which can then be lowered to the optimization dialect (labeled \textit{QuantumSSA}). Code examples for both dialects can be found in Figure~\ref{fig:main}b and Figure~\ref{fig:lowering}. We designed the optimization dialect specifically to enable quantum-classical co-optimization and to enable maximal reuse of compiler components by exposing data dependencies explicitly in the IR. In addition to quantum-specific optimization passes, the IR may thus be optimized using classical transformation passes such as inlining, loop unrolling, and common subexpression elimination (CSE). In a last target-specific step, the optimization dialect may be lowered, e.g., to LLVM IR \cite{llvm04} for simulation, execution on hardware, or resource estimation.

The design of QIRO is guided by the following principles:
\begin{enumerate}
    \item Lowering of existing programming languages into the IR must be simple.
    \item The IR must be capable of supporting state-of-the-art optimization algorithms (quantum and classical).
    \item The IR should enable re-use of existing compilation infrastructure.
\end{enumerate}

We propose two MLIR dialects in order to separate the first requirement from the other two, which are quite different in their nature. As their names suggest, the goal of the input dialect is to enable simple and efficient lowering from existing quantum programming languages, while the optimization dialect is geared toward enabling maximal re-use of existing infrastructure and supporting a wide range of optimizations. Structurally, the two dialects primarily differ in the semantics of how quantum operations interact with qubits.

\label{sec:semantics}
The input dialect represents qubits using \emph{memory-semantics}. That is, qubit allocation returns a unique reference to a qubit. Quantum operations that act on such qubit references do not consume the qubit value, and affect the quantum state via side-effects. Qubit registers function in a similar way, in that allocation returns a unique reference to a register of newly-allocated qubits. An immediate benefit of using memory-semantics is that the structure of the IR inherently prevents a program from violating the no-cloning theorem. As each operation always interacts with the state of the processor via side-effects, there simply is no mechanism available via which a quantum state could be copied. We note, however, that it is not possible to statically guarantee that the same qubit is not passed multiple times to the same quantum operation (and thus aliased), since we allow for quantum register access using dynamic indices. Such cases may be addressed by emitting code that performs this check at run time, e.g. through the RTE.

By contrast, the optimization dialect can be viewed as a quantum-version of SSA that emulates \textit{value-semantics}. By this, we mean that quantum operations consume and return quantum state values instead of operating on qubits via side-effects. These values represent the state of a qubit at a particular time-step in the execution of the quantum program. Note however that such quantum state values are never computed as they would be in a classical setting, instead they merely provide a representation for the purposes of SSA. Using SSA in this way comes with similar benefits to its classical counterpart: (1) the dataflow graph is made explicit in the IR and (2) quantum operations in this dialect are free of side-effects, facilitating optimization (e.g. dead-code elimination for operations whose return values are never consumed).

\section{The Quantum Dialects in Detail}
In this section, we discuss the two proposed dialects in detail, including how to map existing quantum programming languages to the input dialect and how to lower the input dialect to the optimization dialect.

\subsection{Describing Quantum Programs}
In MLIR, all operations are grouped into self-contained units called \emph{modules}, allowing the compiler to process these in parallel. The region of a module is then usually composed of subroutine definitions and external declarations. We distinguish between classical and quantum subroutines. Purely classical subroutines can be placed inside an MLIR \emph{function}, while quantum program segments can be placed inside quantum functions termed \emph{circuits}.

The quantum dialects feature a powerful instruction set with which to interact with the quantum co-processor. This set is composed of all the components for universal quantum computation: qubit initialization into a known state, qubit readout (measurement), and a universal gate set (a set of unitary transformations with which all unitary transformations can be approximated to arbitrary precision).

Additionally, to represent operations at a higher level of abstraction, our IR also features meta-operations (sometimes called functors~\cite{qsharp18}) that modify existing operations in some way. Unless natively supported by the target architecture, these will be lowered via standard or user-defined decomposition routines before execution on the quantum processor.

\subsection{Types}
QIRO defines qubit and quantum register types that are used to represent all quantum data. Higher-level type abstractions, such as integers, fixed-point numbers, and so on, can be implemented on top of these basic types. Each quantum dialect has its own version of the qubit and register types, which reflects the distinction between value- and memory-semantics of the two dialects.
We introduce several additional types to represent quantum operations themselves: a basic type for native single- and two-qubit gates, a type for circuits, and a type for controlled operations (containing the base operation type and number of control qubits). We provide the full list of types in Table~\ref{tab:types}. The dialect shorthands are \code{q.} for the input dialect (Quantum) and \code{qs.} for the optimization dialect (QuantumSSA).

\begin{table}
\caption{Types defined by the quantum dialects.}
\label{tab:types}
\begin{tabular}{l l l}
    \textbf{Type}  & \textbf{Quantum}   & \textbf{QuantumSSA} \\
    \hline
    Qubit          & \code{!q.qubit}    & \code{!qs.qstate} \\
    Qubit Register & \code{!q.qureg<n>} & \code{!qs.rstate<n>} \\
    \hline
    Native 1-Qubit Gate & \multicolumn{2}{l}{\code{!q\textcolor{gray}{s}.u1}} \\
    Native 2-Qubit Gate & \multicolumn{2}{l}{\code{!q\textcolor{gray}{s}.u2}} \\
    Circuit             & \multicolumn{2}{l}{\code{!q\textcolor{gray}{s}.circ}} \\
    Controlled Op       & \multicolumn{2}{l}{\code{!q\textcolor{gray}{s}.cop<n, baseT>}}\\
    \\
    \multicolumn{3}{c}{\small\textbf{n} - register size / number of control qubits}\\
    \multicolumn{3}{c}{\small\textbf{baseT} - type of the underlying operation}
\end{tabular}
\end{table}

\subsection{Operations}
The quantum operations can be broadly separated into four categories: qubit management, native gates, meta-operations, and user-defined operations. An overview can be found in Table~\ref{tab:instruction-set}.

\begin{table*}[t]
\caption{Operations defined by the quantum dialects (dialect prefixes and types omitted).}
\label{tab:instruction-set}
\begin{tabular}{l l l l}
    \textbf{Qubit Management} & \textbf{Native Gates} & \textbf{Meta-Operations} & \textbf{User-defined Operations} \\
    \hline
    \code{\%qb = alloc}                     & \code{H/X/Y/Z/S/T \%q}        & \code{\%op = ctrl \%op, \%q} & \code{circ @name(\%arg..) \{...\}} \\
    \code{\%r~ = allocreg(n)} & \code{R/Rx/Ry/Rz($\phi$) \%q} & \code{\%op = adj \%op}       & \code{call @name(\%arg..)}   \\
    \code{free \%qb}                        & \code{CX \%qb, \%q}           &                              & \code{\%circ = getval @name} \\
    \code{freereg \%r}                      & \code{SWAP \%qb, \%qb}        &                              & \code{apply \%circ(\%arg..)} \\
    \code{\%m = meas \%q}                   & & & \\
    \code{\%qb.., \%r = extract \%r[i..]}   & & & \\
    \code{\%r = combine \%r[i..], \%qb..}   & & & \\
    \\
    \multicolumn{4}{c}{{\small\textbf{\%qb} - qubit value, \textbf{\%r} - register value, \textbf{\%q} - value of either quantum data type}} \\
    \multicolumn{4}{c}{{\small\textbf{\%op} - any quantum op, \textbf{@name} - circuit symbol, \textbf{\%circ} - circuit value}} \\
    \multicolumn{4}{c}{{\small \textbf{n}, \textbf{i} - integers, $\boldsymbol{\phi}$ - floating point, \textbf{\%arg} - any value}}
\end{tabular}
\end{table*}

Qubits and registers are allocated and initialized to the $|0\rangle$ state using the appropriate operations. Allocation errors due to space constraints or other reasons are expected to be handled by the runtime environment. To reinitialize (or reset) a qubit, one can perform a combination of measurement and conditional bit-flip. Note that qubit resources must be explicitly freed, which allows us to enforce that quantum resources may not be used after deallocation, since the freeing operations act as sinks for quantum state values. Measurement operations return a classical bit value or an array thereof, depending on whether the input is a qubit or a register. Note that measurements are performed in the z-basis by default.

The usual single- and two-qubit gates used in the literature are provided as native gate operations and it is straightforward to extend the gate library. All native gates (excluding \code{SWAP}) are overloaded to accept both qubits and registers as their target. The latter can be interpreted as a \textit{foreach} loop, meaning the gate is applied to all qubits inside the given register. Rotation gates additionally accept a continuous angle parameter as either a constant (operation attribute) or a variable (SSA value). A \textit{hermitian} trait is attached to all self-inverse gates, which is used in peephole optimizations.

As the current design of MLIR does not enable operations to directly act on other operations, meta-operations instead act on SSA values \emph{representing} a quantum instruction. These values can be obtained from native gates by omitting the target qubit operands, which must be passed to the meta-operation instead. Multiple meta-operations can also be chained by only supplying target qubit operands to the final one.

Circuits act as ``quantum functions'', in that they form a grouping of operations (quantum and classical) that only have access to values provided as function arguments and those created inside. To enable maximal flexibility, circuits are allowed to accept and return any type of values. There are two ways to invoke a quantum circuit. A direct \emph{call} via the circuit name, and an indirect \emph{application} via a generated circuit value. Indirect circuit application is intended for circuits that have been modified by meta-operations. The code listing below illustrates the use of both constructs.
\vspace{5pt}
\begin{lstlisting}[language=llvm,xleftmargin=0.8cm,xrightmargin=0.8cm,frame=tb,numbers=left,breaklines=false]
q.circ @qft(%r, %n) {
    // define custom QFT operation
}

q.call @qft(%r, %n) // apply QFT

%qft = q.getval @qft
%qft_inv = q.adj %qft
q.apply %qft_inv(%r, %n) // apply invQFT
\end{lstlisting}

\subsection{Meta-Operations}
\label{sec:meta}
Meta-operations require some special care, as they are intended to modify other quantum operations in a way that must be consistent with the laws of quantum mechanics. For native gates, this requirement is always satisfied. However, it is only legal to apply meta-operations to circuits (i.e. user-defined operations) that are also \emph{unitary}. In practice this implies that such circuits must not contain any measurements, and may only contain calls to pure functions, conditions which are asserted on the input IR.

Standard lowering routines for meta-operations on \emph{circuits} can be placed anywhere in the pass-pipeline, such as the one used in the benchmarks of Section~\ref{sec:eval}. They may also enable certain optimization opportunities, as described in Section~\ref{sec:lower-meta}. At the lowest level, adjoint/control decomposition of native gates may be left to the runtime environment, until the final control qubit count or adjoint parity is known.

QIRO also supports custom implementations of adjoint and controlled versions of circuits. These must (1) be marked with a special attribute to identify them as adjoint/control decompositions, and (2) follow the respective naming convention to identify them with the original circuit.

\subsection{Mapping Quantum Languages to the Input Dialect}
\begin{table*}
    \centering
    \resizebox{\textwidth}{!}{
    \begin{tabular}{lllll}
        \textbf{Constructs} & \textbf{Q\#} & \textbf{Qiskit} & \textbf{Silq} & \textbf{QIRO} \\
        \hline
        Allocation &                                  & \code{resArr = ClassicalRegister(n)} & & \\
                   & \code{using (q = Qubit()) \{\}}  &                                      & \code{q := 0:$\mathbb{B}$} & \code{\%q = q.alloc -> !q.qubit} \\
                   & \code{using (r = Qubit[n]) \{\}} & \code{r = QuantumRegister(n)}        & \code{r := array(n,0:$\mathbb{B}$):$\mathbb{B}$[]}      & \code{\%r = q.allocreg(n) -> !q.qureg<n>} \\
                   &                                  & \code{c = QuantumCircuit(r, resArr)} & & \\
        Deallocation & \multicolumn{1}{c}{automatic} & \multicolumn{1}{c}{automatic} & \multicolumn{1}{c}{automatic} & \code{q.free \%q : !q.qubit} \\
                     &                               &                               & & \code{q.freereg \%r : !q.qureg<n>} \\
        \hline
        Measurement & \code{let res = M(q)}         & \code{c.measure(r[i], resArr[i])} & \code{res := measure(q)} & \code{\%res = q.meas \%q : !q.qubit -> i1} \\
                    & \code{let resArr = MultiM(r)} & \code{c.measure(r, resArr)}       & \code{resArr := measure(r)} & \code{\%resArr = q.meas \%r : !q.qureg<n> -> memref<nxi1>} \\
        \hline
        Native gates & \code{H(q)}            & \code{c.h(r[i])}            & \code{q := H(q)}             & \code{q.H \%q : !q.qubit} \\
                     & \code{Rz($\phi$, q)}   & \code{c.rz($\phi$, r[i])}   & \code{q := rotZ($\phi$,q)}   & \code{q.Rz($\phi$) \%q : !q.qubit} \\
                     & \code{CNOT(qc, qt)}    & \code{c.cx(r[i], r[j])}     & \multicolumn{1}{c}{--}       & \code{q.CX \%qc, \%qt : !q.qubit, !q.qubit} \\
        \hline
        User operations & \code{function Foo(args..) : resT \{\}}  & \code{def Foo(args..):}            & \code{def Foo(args..) \{\}} & \code{func @Foo(args..) -> !resT \{\}} \\
                        & \code{let ret = Foo(args..)}             & \code{ret = Foo(args..)}           & \code{ret := Foo(args..)}   & \code{\%ret = call @Foo(args..) : (!argT) -> !resT} \\
                        \\
                        & \code{operation Bar(args..) : Unit \{\}} & \code{bar = QuantumCircuit(m)}     & \code{def Bar(args..) \{\}} & \code{q.circ @Bar(args..) -> !resT \{\}} \\
                        & \code{Bar(args..)}                       & \code{Bar = bar.to\_instruction()} & \code{args.. = Bar(args..)} & \code{q.call @Bar(args..) : !argT -> !resT} \\
                        &                                          & \code{c.append(Bar, r[0, m])}      & & \\
        \hline
        Meta-operations & \code{Controlled X(qc, qt)} & \code{CX = XGate().control()}      & \code{if qc \{}             & \code{\%X = q.X -> !q.gate} \\
                        &                             & \code{c.append(CX, [r[i], r[j]])}  & \code{ \ qt := X(qt)}        & \code{q.ctrl \%X, \%qc, \%qt : !q.gate, !q.qubit, !q.qubit} \\
                        &                             &                                    & \code{\}}                   & \\
                        & \code{Adjoint X(q)}         & \code{Xdg = XGate().inverse()}     & \code{q := reverse(X)(q)}   & \code{q.adj \%X, \%qt : !q.gate, !q.qubit} \\
                        &                             & \code{c.append(Xdg, r[i])}         &                             & \\
                        \\
                        & \code{Adjoint Bar(q)}       & \code{Bar = bar.to\_instruction()} & \code{q := reverse(Bar)(q)} & \code{\%Bar = q.getval @Bar -> !q.circ} \\
                        &                             & \code{BarA = Bar.inverse()}        &                             & \code{\%BarA = q.adj \%Bar : !q.circ -> !q.circ} \\
                        &                             & \code{c.append(BarA, r[0, m])}     &                             & \code{q.apply \%BarA(\%q) : !q.circ(!q.qubit)} \\
        \hline
        Conditionals & \code{if (res = One) \{} & \code{c.x(r[i]).c\_if(resArr[j], 1)} & \code{if res \{}    & \code{scf.if \%res \{} \\
                     & \code{ \ X(q)}           &                                      & \code{ \ q := X(q)} & \code{ \ q.X(\%q) : !q.qubit} \\
                     & \code{\}}                &                                      & \code{\}}           & \code{\}}\\
        \hline
        Loops & \code{for (i in 0 .. Length(r)) \{\}} & \code{for i in range(0, len(r)):} & \code{for i in [0, n) \{\}} & \code{affine.for \%i = 0 to \%n \{\}} \\
              &                                       &                                   &                             & \code{scf.for \%i = \%c0 to \%n step \%c1 \{\}} \\ 
              &                                       &                                   & \multicolumn{2}{l}{\ \ \ \ \ \ - - - - \textit{register size must be kept around as a value} - - - -} \\
              \\
              & \code{repeat \{}                      & \multicolumn{1}{c}{--}            & \code{while res \{}         & \code{\textasciicircum repeat:} \\
              & \code{ ...}                           &                                   & \code{ \ ...}               & \code{ \ ...} \\
              & \code{ let res = M(q)}                &                                   & \code{ \ res = measure(q)}  & \code{ \ \%cond = q.meas \%q : !q.qubit -> i1} \\
              & \code{\} until (res == Zero)}         &                                   & \code{\}}                   & \code{ \ cond\_br \%cond, \textasciicircum repeat, \textasciicircum next} \\
        \hline
    \end{tabular}}
    \caption{Quantum programming constructs and their representation in high-level languages and QIRO. Automatic here refers to scope-based deallocation. As a special case in Qiskit, all operations must be called from a circuit object (e.g. \code{c.h()}). This object-oriented notation is not to be confused with dialect prefixes in QIRO (\code{q.} and \code{qs.}).}
    \label{tab:lang_map}
\end{table*}

Since QIRO is capable of representing languages based on the widely used circuit model of quantum computing, many such languages are able to benefit from performant static code optimization enabled by our design. In Table~\ref{tab:lang_map} we show how high- and low-level constructs map to our IR from a selection of languages with different foci, namely Q\# (high-level quantum and classical code), Qiskit (NISQ/eDSL), and Silq (intuitive algorithm development).

Qubit and register allocation, measurement, and low-level quantum gates all map to our IR in a straightforward fashion. Higher-level quantum operations should be translated to circuits, whereas functions are reserved for purely classical code. An advantage of MLIR is the ability to express both conditional expressions and for loops in the form of structured control flow, using operation nesting rather than the flat block structure traditionally used in SSA. More complex control flow can be represented using blocks, such as the while or repeat until success loops shown in Table~\ref{tab:lang_map}. Whenever possible, for loops should be mapped to the affine dialect to take advantage of the dialect's powerful optimization passes. For further information we refer to Appendix~\ref{app:mapping}, where we discuss how Q\# constructs may be mapped to our IR in more detail.

\begin{figure}
    \begin{subfigure}{\linewidth}
        \centering
        \includegraphics[width=.95\linewidth]{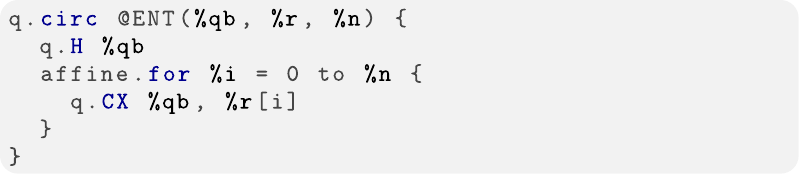}
    \end{subfigure}
    $\mathbf{\downarrow}$
    \begin{subfigure}{\linewidth}
        \centering
        \includegraphics[width=.95\linewidth]{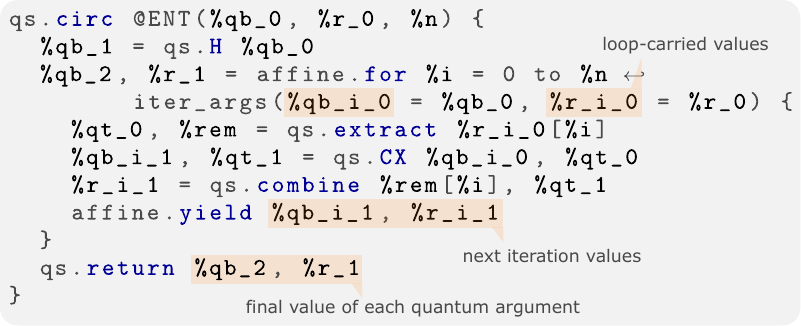}
    \end{subfigure}
    \caption{An entanglement circuit is lowered from the input dialect to the optimization dialect. The qubit \code{\%qb} is entangled with a register \code{\%r} of size \code{\%n} using an affine loop. In the optimization dialect, loop structures make use of loop-carried values to enable value-semantics.}
    \label{fig:lowering}
\end{figure}

\subsection{Lowering to the Optimization Dialect}
The input dialect is lowered to the optimization dialect by recursively traversing the nested structure inside a module in a post-order fashion. This ensures that nested operations are replaced first, since these need to be available when constructing the new parent operation. The main purpose of the lowering pass is to transform from memory-semantics into value-semantics, as well as to transform some convenience features into explicit code. Figure~\ref{fig:lowering} depicts a small example program lowered to the optimization dialect.

\paragraph{Gates.} In the input dialect, gates act on qubit references implicitly via side-effects. To convert this behaviour to value-semantics, a gate operation needs to generate a new state value for each unique quantum argument. During lowering, the transformation pass keeps an updated map of the most recent state value for each qubit reference. Gate arguments are then replaced with the latest state value, and the map is updated with the return values of the quantum operation.

\paragraph{Circuits.} Similar to MLIR functions, circuits are isolated portions of code that can only reference those SSA values defined in the body and argument list. Thus, the lowering pass can keep a local map for qubit state values when entering a circuit, which is initially populated with the circuit's arguments. The signature must also be updated to include the return type of each quantum argument. Block terminators that transfer control flow back out of the circuit (often implicit in the input dialect) are replaced with a \code{return} operation that returns all values present in the qubit state map when reaching the end of the circuit (see Figure~\ref{fig:lowering}).

\paragraph{Loops.} \code{For} loops from both the \emph{structured control flow} (SCF) and \emph{affine} dialects are lowered to value-semantics by exploiting loop-carried values for quantum states, as shown in Figure~\ref{fig:lowering}. The \code{iter\_args} parameter defines these loop-carried values and provides their initial values. Closing off the loop body, the \code{yield} operation returns the values to be passed to the next iteration, or to be returned upon reaching the final iteration.

\paragraph{Register access.} At its core, SSA form is best suited to represent dataflow of \emph{scalar} variables. When dealing with aggregate structures such as arrays (qubit registers) and global memory, it has traditionally been difficult to directly represent these via SSA~\cite{aho86}. Consequently, memory is frequently modeled and represented separately, such as in LLVM and MLIR. Memory references can be used to \code{load} memory elements into scalar SSA values, which can be easily operated on, and eventually \code{stored} back into memory.

This problem is also side-stepped in the input dialect, which conveniently represents qubit register access via indices whenever the register value is used. This preserves the uniqueness of qubit and register references required by the input dialect. However, it does obscure data dependencies between individual qubits of the register in the dataflow graph. To enable optimizations requiring single qubit dataflow, we make these data dependencies \emph{locally} explicit in the optimization dialect. Similar to the memory load/store model, individual qubits or qubit slices are \code{extracted} from registers and re-\code{combined} after use. Whenever possible, we use static data analysis to consolidate such extract/combine operations in order to directly expose the dataflow of register elements. In this way, we can exploit proper SSA semantics inside optimization passes, without needing to handle register dataflow analysis for every optimization. This is in line with our design principle of modularizing reusable components.

\section{Transformations \& Analyses}

QIRO supports and facilitates the implementation of many transformations and analyses important to compilation of mixed quantum-classical programs. In particular, static optimization passes are much better suited for large-scale quantum program optimization, in contrast to NISQ-focused, run-time optimization systems, as we show in Section~\ref{sec:eval}. Furthermore, quantum resource estimation is a compute-intensive analysis that can profit from our proposed IR and compilation stack.

\begin{figure}
    \includegraphics[width=0.95\linewidth]{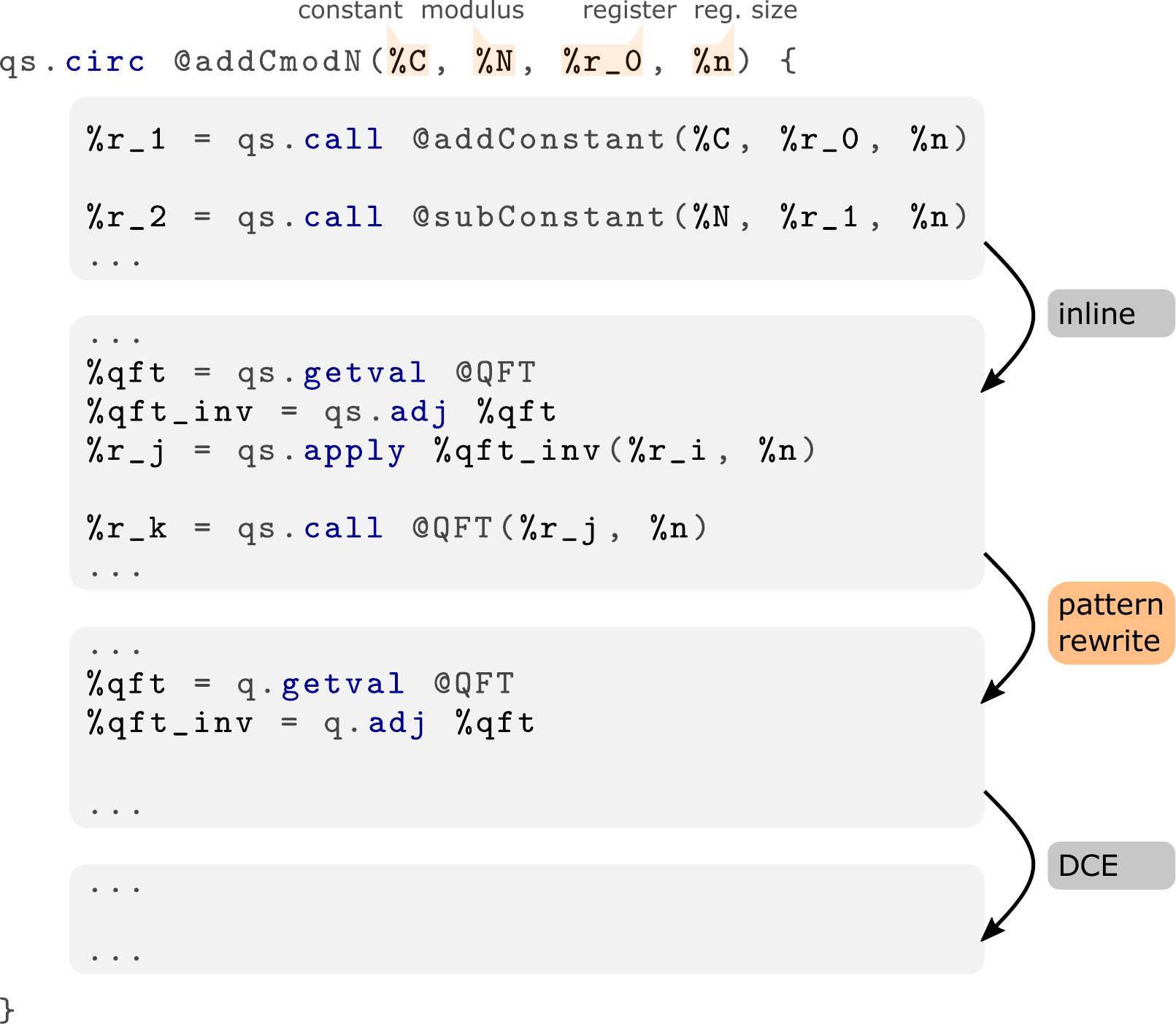}
    \caption{Application example showing different passes interacting with each other to remove superfluous computations. This modular adder routine makes successive calls to constant-addition routines which each begin and end by invoking the QFT and QFT$^\dagger$ operations, respectively.}
    \label{fig:passes}
\end{figure}
    
\subsection{Classical Optimizations}
MLIR provides us with a variety of useful transformations applicable to a program written in our IR, maximizing reuse of compiler components where appropriate. The following MLIR passes can be used out-of-the-box on mixed classical-quantum programs, with the exception of \emph{Inlining} which was slightly modified to work on quantum callables. \footnote{While the list provided here focuses on classical passes that affect the \emph{quantum} program parts, purely classical program sections can naturally be optimized with the usual passes both at the IR-level within MLIR, and at the backend-level e.g. within LLVM.}

\paragraph{Canonicalization} While this technically refers to bringing code into a single ``canonical'' form, this pass includes many important optimizations such as \textit{constant folding} and \textit{dead code elimination} (DCE). Note that DCE not only applies to classical operations, but also directly to side-effect free quantum operations whose return values go unused (such as in Figure~\ref{fig:passes}) due to the SSA structure of the optimization dialect.

\paragraph{CSE} This pass can eliminate duplicate (classical) expressions and replace them with a single value. Similar to the canonicalizer, this can improve knowledge of dataflow inside registers, similar to what is demonstrated in Figure~\ref{fig:main}b.

\paragraph{Inlining} Traditional function inling is generally useful for speeding up the computation of small functions, trading program size for speed. Moreover, this pass was slightly adapted to inline quantum circuits, which is often vital to expose optimization opportunities. As the instruction sequences inside circuits are often pre-optimized by hand, many more opportunities arise once multiple circuit calls are merged together (as demonstrated in Figure~\ref{fig:passes}).

\paragraph{Affine loop unrolling} Loops in a quantum program expressed via the affine dialect can be unrolled (full or by a factor) using this pass. In the same vein as the inlining pass, unrolling allows for the optimization of quantum gates on the boundary of loop iterations. One should note that the loop boundary optimization described below is an alternative to using affine loop unrolling which can be applied to non-affine loops as well.

\subsection{Quantum Circuit Optimizations}
It is clear that traditional approaches to \emph{circuit} optimization can be supported by running optimizers on sections consisting only of quantum operations, say e.g. a loop body. We note, however, that the interaction between quantum and classical optimizations, such as peephole optimization and loop-unrolling, may further increase the impact on the quantum resource requirements. Moreover, many identity-based heuristics found in quantum circuit optimization papers~\cite{nam18} are naturally suited for implementation using MLIR's pattern rewriter, as these represent simple DAG-to-DAG transformations. Whereas the classical optimizations given above are already available in MLIR, the quantum optimization passes below were re-implemented in QIRO to demonstrate its capability to support optimization passes.

\subsubsection{Local Register Dataflow}
This pass is not based on existing optimizations, instead being devised for the specifics of QIRO's design. As a precursor to other optimizations, we wish to consolidate \code{extract} / \code{combine} operations that occur when translating register accesses from the input dialect to the optimization dialect (see Figure~\ref{fig:main}b). To enable optimizations in these scenarios, we introduce a transformation pass that merges \code{combine} instructions with subsequent \code{extracts} whenever they are linked in the use-def chains and satisfy the following restrictions: If there exists overlapping qubit indices, these can be removed from both operations extending the qubit values lifetime from the first block into the second. If all indices are distinct, we delay re-combining qubits from the first block until the end of the second block, merging the two \code{combine} operations. With static indices, these assertions can always be made. With dynamic indices, on the other hand, we perform this optimization only if dataflow analysis on the indices can guarantee one of the conditions. Additionally, \code{combine}--\code{combine}, \code{extract}--\code{extract}, and empty \code{extract}--\code{combine} patterns are optimized in a similar fashion to expand the regions of locally available register dataflow.

\subsubsection{Peephole Optimizations.}
Simple peephole (or window) optimizations are straightforward to implement on the optimization dialect using MLIR's pattern rewriting framework. Unitary gate cancellation is among the most common ones, where we differentiate between two cases: two identical but Hermitian (or self-inverse) gates, and a general gate followed by its unitary inverse. Additionally, a quantum analog to classical constant folding, namely merging parametrized rotation gates, also falls in this category.

\paragraph{Hermitian Operations} Matching is performed on all quantum operations carrying the Hermitian trait. As every quantum data type in the input must also be present in the output, we follow the use-def chain of each quantum operand value to its defining operation. If the operation is the same for all such operands, is the same \emph{kind} as the operation we started with, and all remaining arguments are identical, we have found a match. Replacing the matched pattern is then as simple as replacing all uses of the returned values from operation 2 by the input values of operation 1, and erasing both operations.
\medskip
\begin{center}
\begin{minipage}{4.7cm}
\begin{lstlisting}[language=llvm,frame=tb,numbers=left,xleftmargin=0.5cm,breaklines=false]
%a1, %b1 = qs.CX %a0, %b0
%a2, %b2 = qs.CX %a1, %b1
%a3 = qs.H %a2
\end{lstlisting}
\end{minipage}
$~\longrightarrow~$
\begin{minipage}{2.9cm}
\begin{lstlisting}[language=llvm,frame=tb,numbers=left,xleftmargin=0.5cm,breaklines=false]


%a3 = qs.H %a0
\end{lstlisting}
\end{minipage}
\end{center}

\paragraph{General Operations} In the more general case, we must extend the matching over the \code{adjoint} meta-operation, but otherwise remains largely the same. Starting from either the unitary gate or the adjoint op, we trace back all quantum data operands for a match, and additionally follow the unitary operand of the adjoint op to its definition to determine if the operation kind matches.

The replacement step consists of erasing the matched adjoint and unitary gate operations, along with replacing all uses of the quantum data operands. We note that this pattern may be extended straightforwardly from native operations to user-defined quantum circuits.
\medskip
\begin{center}
\begin{minipage}{4cm}
\begin{lstlisting}[language=llvm,frame=tb,numbers=left,xleftmargin=0.5cm,breaklines=false]
%a1 = qs.T %a0
%t = qs.T
%a2 = qs.adj %t, %a1
%a3 = qs.H %a2
\end{lstlisting}
\end{minipage}
$~\longrightarrow~$
\begin{minipage}{3cm}
\begin{lstlisting}[language=llvm,frame=tb,numbers=left,xleftmargin=0.5cm,breaklines=false]

%t = qs.T

%a3 = qs.H %a0
\end{lstlisting}
\end{minipage}
\end{center}

\paragraph{Merging of Rotations} Again we follow a similar procedure for matching two adjacent (in this case rotation) gates on the use-def graph, this time replacing them with a single operation that has as parameter the sum of the individual rotation angles. The new rotation angle is computed by a classical operation inserted in front of the rotation, or directly inserted into the operation in case of static arguments.
\smallskip
\begin{center}
\begin{minipage}{3.8cm}
\begin{lstlisting}[language=llvm,frame=tb,numbers=left,xleftmargin=0.45cm,breaklines=false]
%a1 = qs.Rz(0.1) %a0
%a2 = qs.Rz(0.3) %a1
%a3 = qs.H %a2
\end{lstlisting}
\end{minipage}
$~\longrightarrow~$
\begin{minipage}{3.8cm}
\begin{lstlisting}[language=llvm,frame=tb,numbers=left,xleftmargin=0.45cm,breaklines=false]
%a1 = qs.Rz(0.4) %a0

%a3 = qs.H %a1
\end{lstlisting}
\end{minipage}
\end{center}

\subsubsection{Loop Boundary Optimization.}
This optimization analyzes loop structures across their iteration boundaries, which is particularly effective on loop bodies with a symmetric structure, such as compute/uncompute segments. Repeating such a body multiple times leads to many redundant instructions that undo and redo the same operation at the end and beginning of every iteration. The optimization is performed by walking use-def chains from both ends of the loop body and keeping track of matches, similar to previously described peephole optimizations. Operations that cancel can be hoisted out of the loop body and placed right before and after the loop operation (for the first and last iteration). A similar optimization is possible for rotations, which can be merged across loop iterations.
\medskip
\begin{center}
\begin{minipage}{8.3cm}
\begin{lstlisting}[language=llvm,frame=tb,numbers=left,xleftmargin=0.5cm,breaklines=false]
%a1 = scf.for %i=0 to 6 iterargs(%a_0 = %a) {
  %a_1 = qs.H %a_0
  %a_2 = qs.T %a_1
  %a_3 = qs.H %a_2
  yield %a_3
}
\end{lstlisting}
\end{minipage}
$\mathbf{\downarrow}$
\begin{minipage}{8.3cm}
\begin{lstlisting}[language=llvm,frame=tb,numbers=left,xleftmargin=0.5cm,breaklines=false,aboveskip=0.3cm]
%a0 = qs.H %a
%a1 = scf.for %i=0 to 6 iterargs(%a_0 = %a0) {
  %a_2 = qs.T %a_0
  yield %a_2
}
%a2 = qs.H %a1
\end{lstlisting}
\end{minipage}
\end{center}

\subsection{Partial Lowering \& Decompositions}
In the process of translating a quantum program to executable code, it is necessary to \emph{decompose} complex quantum operations into simpler gates supported by the instruction set of the target architecture. At the lowest level, this is best left to an architecture-aware backend, but, where sensible, one should aim to perform these decompositions within the IR so as to leverage the multi-level rewrite infrastructure. These decompositions can then be interleaved with optimization passes to maximize optimization potential. For example, adjoint and controlled versions of user-defined operations are fully expressible within the IR, and can benefit from optimizations pre- and post-decomposition.

\label{sec:lower-meta}
\paragraph{Adjoint-Circuit Lowering}
A standard adjoint lowering pass on \textit{unitary} circuits can be implemented by generating a new circuit in which the order of quantum operations has been reversed, and the adjoint meta-operation is applied to each of the operations inside. An optimization opportunity arises here for native gates with the \code{hermitian} trait, as such gates are self-inverse and need not be modified with an adjoint operation.

\paragraph{Controlled-Circuit Lowering}
Similarly, controlled \textit{unitary} circuits can be lowered by generating a new circuit and propagating the control meta-operations to each quantum operation inside. The qubits and registers upon which the circuit is controlled must be passed as new arguments to the generated circuit. Another optimization opportunity arises here when using attributes, provided by the frontend, to mark special \code{compute}/\code{uncompute} sections in the code~\cite{haner18}. This allows omitting the control propagation on these sections, while still producing the same computation.

\subsection{Resource Estimation}
\label{sec:res-est}
\begin{table*}[b]
\caption{Pass sequence used within QIRO to run optimizations and perform resource estimation on Shor's algorithm.}
\label{tab:passes}
\begin{tabular}{p{0.22\linewidth}|p{0.73\linewidth}}
    \textbf{Pass}               & \textbf{Description} \\
    \hline
    --convert-mem-to-val        & Converts from the input to the optimization dialect, switching qubit semantics in the process. \\
    --lower-ctrl                & Converts controlled circuit to new circuit with controls propagated to operations inside, does not propagate on compute/uncompute sections indicated by \code{compute}/\code{uncompute} attribute. \\
    --strip-circ                & Removes unused circuit definitions. \\
    --canonicalize              & Built-in MLIR canonicalization pass, also performs DCE, constant folding, and operation rewrite patters, including the \emph{local register dataflow} patters. \\
    --strip-circ                & Removes unused circuit definitions. \\
    --circuit-inline            & Inlines circuit calls with the corresponding circuit body (except where \code{no\_inline} / \code{no\_inline\_target} attribute present). \\
    --strip-circ                & Removes unused circuit definitions. \\
    --canonicalize              & Built-in MLIR canonicalization pass, also performs DCE, constant folding, and operation rewrite patters, including the \emph{local register dataflow} patters. \\
    --strip-circ                & Removes unused circuit definitions \\
    --quantum-gate-opt          & Performs hermitian gate cancellation, adjoint gate cancellation, adjoint circuit cancellation, rotation gate folding, controlled-rotation gate folding, and loop-boundary optimization. \\
    --canonicalize              & Built-in MLIR canonicalization pass, also performs DCE, constant folding, and operation rewrite patters, including the \emph{local register dataflow} patters. \\
    --count-resources           & Replaces quantum gates with resource counters. \\
    --convert-scf-to-std        & Built-in MLIR conversion from structured control flow to standard dialect. \\
    --convert-vector-to-llvm    & Built-in MLIR conversion from vector to llvm dialect. (Final resource counts are printed using the \code{vector.print} operation.) \\
    --convert-std-to-llvm       & Built-in MLIR conversion from standard to llvm dialect.
\end{tabular}
\end{table*}

One particularly efficient way to generate \emph{resource} (or quantum gate) counts, is to strategically lower and replace quantum operations by classical ones in a way that preserves the structure of the quantum program, and increments simple counters for each measured resource~\cite{meuli2020}. In this process, all adjoint and controlled circuits must first be lowered using the passes described above. Then, for each native gate, a formula can be provided to indicate the decomposition cost of their controlled and adjoint versions in terms of the tracked resources. Thus, all native gate invocations can directly be replaced with classically computed counter increments. 

Any remaining and unused operations from the quantum dialects must be stripped, and circuit definitions and calls must be converted to standard MLIR functions. Special care must be taken when removing measurements, so as to provide a conservative estimate for computations that depend on measurement outcomes~\cite{meuli2020}. Finally, the purely classical IR can be lowered to LLVM IR using standard MLIR infrastructure, and subsequently compiled into an executable outputting the final resource counts for a given input size.
A simplified implementation of this resource estimator was used to compute the rotation gate counts in Section~\ref{sec:bench-resource}. We note that the run time for computing resource estimates may be reduced further using custom compiler passes such as the ones employed by \citet{meuli2020}.

\subsection{Run-time Optimization}
We stress that while it is beneficial to run optimizations at compile time, certain optimizations may provide additional benefits at run time, and should be seen as complementary to our work. This is especially important for quantum programs where the quantum circuits being executed on the QPU depend heavily on run-time parameters (user-input as well as qubit measurements). In such cases, run-time optimizations may further reduce resource requirements. Additionally, mapping of a quantum program to specific quantum computer architectures often introduces new optimization opportunities as well, due to transformations required by varying qubit layouts and connectivities, as well as differing native gate sets.

\section{Evaluation}
\label{sec:eval}
We evaluate the performance of our IR for optimization and resource estimation on the example of Shor's algorithm. Similarly, we evaluate the effectiveness of standard optimizations performed at compile time compared to those performed at run time on the same algorithm. Factoring large numbers is anticipated to be one of the earlier applications of large-scale quantum computers with proven exponential speed-up. With the number of elementary gates growing quickly with the input size, compilation systems that rely on building up large circuit data structures in Python (such as ProjectQ and Qiskit) are slow at optimization and estimating resource requirements. With the projected need for error-corrected computation, non-Clifford gates such as general rotation gates are expected to make up the bulk of computation time. We thus focus on reporting the number of single-qubit rotations during our evaluation. Results from our prototype implementation are labeled QIRO.

Our evaluation uses an implementation of Shor's algorithm by \citet{beauregard03}, excluding manual optimizations such as canceling the (inverse) QFTs of subsequent Fourier adders~\cite{draper2000}, resulting in an identical implementation to the one in ProjectQ~\cite{projectq18}. We implemented the algorithm directly in QIRO's input dialect as a starting point for our compilation pipeline. This implementation can be found in full in Appendix~\ref{app:shors}. The program is then subjected to the following steps:
\begin{itemize}
\item {input-to-optimization dialect lowering}
\item {program optimization}
\item {resource estimation conversion}
\item {translation to LLVM IR}
\item {LLVM compilation + linking}
\end{itemize}

During this process, we collect three different benchmark metrics: the compilation time (everything up to the execution step), the execution time, and the resource estimates reported by the generated executable. Note that this executable is a classical one, stripped of all quantum instructions for the purpose of resource estimation as described in Section~\ref{sec:res-est}. We thus evaluate how quickly our framework is able to perform optimizations (compilation time), how quickly it can provide resource estimates (compilation + execution time), and how effective its static optimizations are (reported resource requirements). Our results are then compared to those obtained in ProjectQ and Qiskit, where the optimization and resource estimation process used is the standard one for the respective framework, with similar optimization levels. We note that Q\# does not currently offer quantum program optimizations and is thus unsuitable for this comparison.

\subsection{Experimental Setup}
Benchmarking was performed on an Intel Core i7-7700HQ @ 3.5GHz running Windows 10 build 18363. The software packages used include: Python 3.7.6, ProjectQ 0.5.1, Qiskit 0.23.1, LLVM/Clang 10.0.0, and MLIR built from source from the master branch dated 2020/10/25. Program transformation at the MLIR level is handled by an adapted version of the modular MLIR optimizer (mlir-opt). Further tools are used in the translation from MLIR to LLVM IR (mlir-translate), the compilation of LLVM IR by the LLVM static compiler (llc), and the linking phase (clang). Table~\ref{tab:passes} shows the built-in and custom QIRO passes used in the benchmark. Execution was timed using the \code{hyperfine}\footnote{\url{https://github.com/sharkdp/hyperfine}} command-line tool, using the median and standard deviation from each sample set.

\subsection{Optimization \& Resource Estimation Run Time}
A comparison of the time it takes to optimize and obtain resource counts for Shor's algorithm is shown in Figure~\ref{fig:benchmark}. The numbers to factor were chosen as $N=2^n-1$, where $n\in\{2,...,64\}$ is the number of bits. Inputs beyond 64-bit integers are also possible but would require to adapt our implementation to larger fixed-precision or infinite-precision arithmetic. As shown, compilation in QIRO including all implemented optimizations took a mere 551$\pm$4 ms, independent of $n$. Moreover, the run time of the resource estimation executable on the entire input range is situated between 10.2$\pm$0.4 ms to 560$\pm$20 ms. By contrast, inputs to ProjectQ were limited to $n=8$ due to execution times (optimization + resource estimation) reaching 162$\pm$3 s, an almost 300-fold increase over QIRO's total running time on the same input (0.561$\pm$0.004 s). Similarly, Qiskit reaches 74.4$\pm$0.1 s on 11 qubits for optimization and resource estimation, a 130-fold increase over QIRO (0.562$\pm$0.004 s).

While some of the difference in the measured run times may be attributed to different implementation details (e.g. loop-heavy operations in Python compared to MLIR's C++ implementation), a remarkable distinction of conceptual nature is the fact that the run time of optimizations in QIRO is independent of program inputs, and thus inherently more efficient. This can be attributed to the fact that QIRO is able to perform optimizations at compile time, while the two other frameworks require program input propagation and control flow resolution before performing optimizations.

We note that true application-scale inputs for breaking RSA are expected to be $n\sim 2000$. Extrapolating our measurements to 2000 bits based on a polynomial of degree 4 (as the number of gates scales quartically), we roughly estimate to be able to process Shor's algorithm in QIRO on the order of a week, rather than upwards of $10^4$ years in ProjectQ and $10^3$ years in Qiskit. To further speed up resource estimation, we plan to implement custom compiler passes such as the ones proposed by \citet{meuli2020}.

\subsection{Effectiveness of Static Optimizations}
\label{sec:bench-resource}
For this benchmark, we implemented a series of simple peephole optimizations and measure their effectiveness in QIRO compared to ProjectQ. Our focus is not on the effectiveness of the optimizations themselves, but rather to investigate the difference in effectiveness when the same optimizations are performed at compile time rather than run time. The implemented optimizations include Hermitian gate cancellation on the native gate set, generalized adjoint cancellation via meta-operations (including on user-defined circuits), successive rotation gate merging, and loop boundary optimization. As shown in Figure~\ref{fig:resources}, practically all ($\sim$99.8\% at $n=8$) optimization opportunities on rotation gates exploited by ProjectQ are also identified by our optimizer at compile time, not least due to our ability to statically optimize across loop boundaries, without which only $\sim69.4\%$ would be identified. Our approach comes with the additional benefit that the optimizer's execution time is independent of a program's input size, making it especially promising for optimization of very large-scale quantum programs.

\begin{figure}
    \centering
    \includegraphics[width=\linewidth]{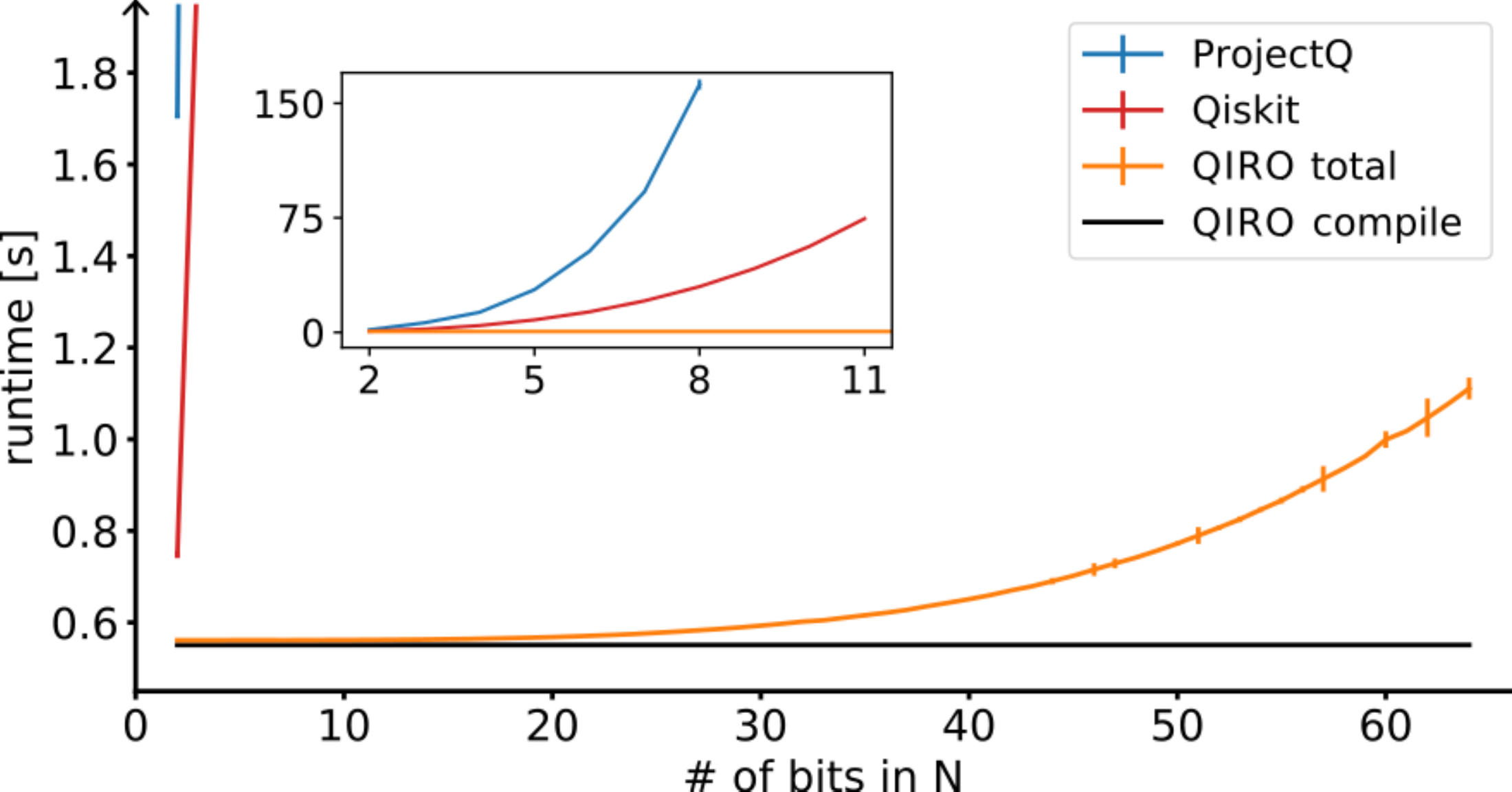}
    \caption{Execution time to process Shor's algorithm by different frameworks and output resource counts. ProjectQ and Qiskit perform circuit optimizations at run time, whereas QIRO performs these exclusively at compile time. The number to factor is chosen as $N=2^n-1$ with $n \in [2,64]$. The insert is a scaled version to show large measurement values.}
    \label{fig:benchmark}
\end{figure}

\begin{figure}
    \centering
    \includegraphics[width=\linewidth]{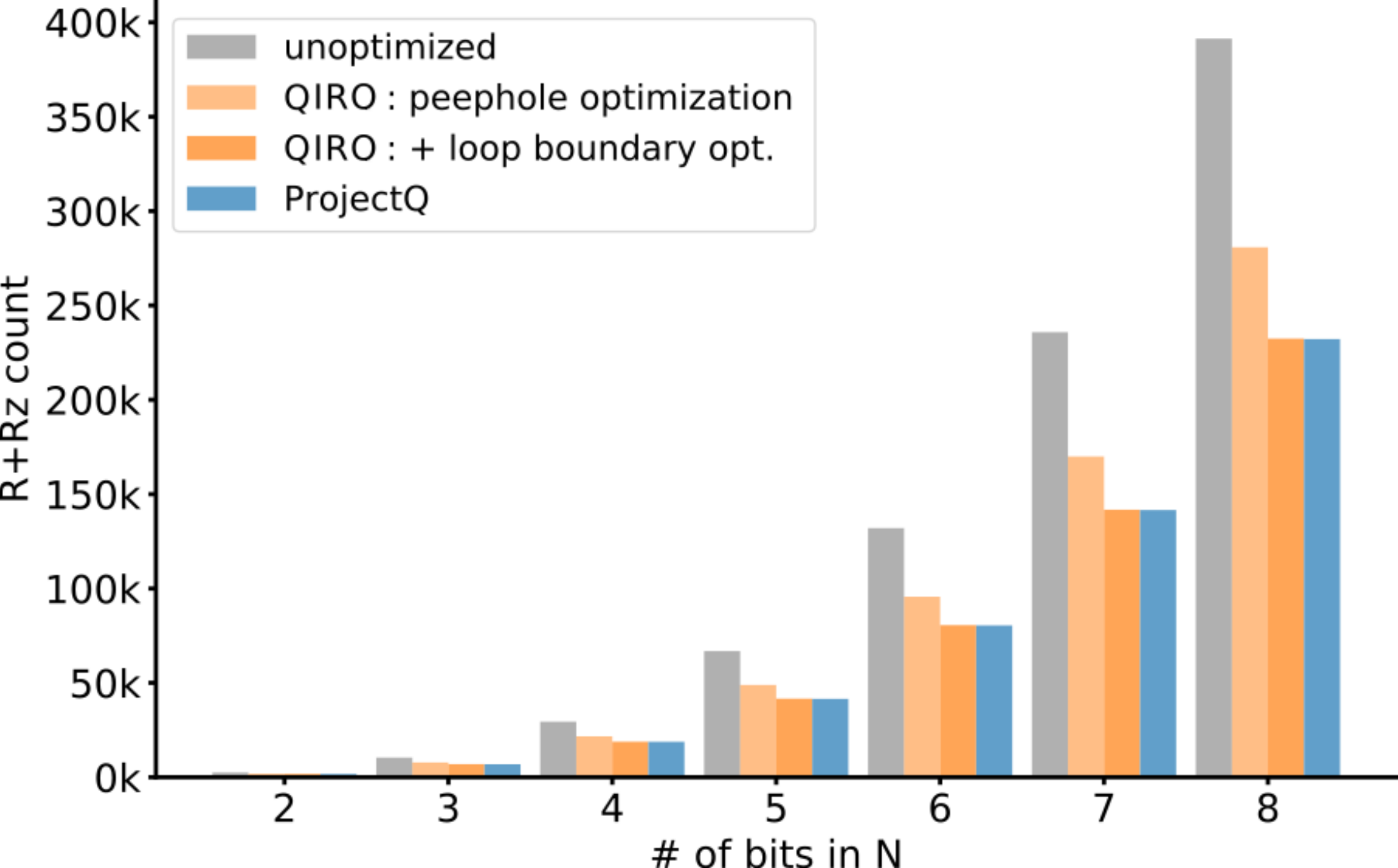}
    \caption{Fraction of optimization opportunities identified by QIRO at compile time. Rotation gate counts for Shor's algorithm factoring $N=2^n-1$, with $n \in [2,8]$, are compared to those obtained in ProjectQ using run-time optimizations.}
    \label{fig:resources}
\end{figure}

\section{Related Work}
\label{sec:related}

\subsection{Intermediate Representations}
To the best of our knowledge, the results presented by \citet{mccaskey2021} have been the only other effort to leverage the MLIR framework for quantum compilation. We see their work as complementary to ours, as it strictly focuses on translation aspects of compiling quantum programs down from QASM to the LLVM-based QIR, without considering the MLIR-based representation as a platform for quantum program optimization.

The recently introduced Quantum Intermediate Representation (QIR)~\cite{qir2020} is another IR specifically crafted as a language- and hardware-agnostic intermediate representation for integrated classical-quantum programs. QIR is a set of specifications for representing quantum programs in the LLVM IR, with the goal of leveraging the performant LLVM compiler infrastructure. However, we note that the employed memory-semantics for quantum operations limits reuse of optimization passes that rely on dataflow analysis -- a problem that we address by introducing a separate optimization dialect in MLIR.

LLVM IR has also been used in the ScaffCC compiler~\cite{scaffcc14} for the C-based Scaffold programming language~\cite{scaffold12}. However, Scaffold programs are restricted to descriptions of fully specified circuits, i.e., all classical control flow present in the input must be statically resolvable to produce flattened circuits. This constitutes a limitation that is not present in our work.

Most existing quantum programming languages represent quantum circuits as gate lists or DAGs, e.g., Qiskit~\cite{qiskit19}, ProjectQ~\cite{projectq18}, pyQuil~\cite{quil16}, and Cirq~\cite{cirq2020}. As a result, static co-optimization of quantum-classical programs with nontrivial control flow is infeasible. As a remedy, these frameworks usually employ run-time optimizations. However, the execution time of optimizations then scales with problem size, as can be seen in Figure~\ref{fig:benchmark}. This makes these approaches ill-suited for quantum program optimization at application scale.

\subsection{Quantum Program Optimization}
In addition to quantum analogs of classical optimizations (e.g. constant-folding at different levels of abstraction~\cite{haner18}), there exists a host of quantum-specific optimizations that are mostly targeted at quantum circuits: Circuit synthesis may be employed to re-synthesize small subcircuits~\cite{forest15,paetznick13,iten16,amy13}. Furthermore, optimization using phase polynomials has proven to be effective~\cite{amy14}, especially when combined with other heuristics to tackle larger universal circuits~\cite{nam18}. Moreover, assertion-based optimization has been proposed to optimize quantum programs at higher levels of abstraction~\cite{haner2020optimization}.

All of these optimization algorithms may be integrated into QIRO as transformation passes. Indeed, we have implemented a generalization of constant-folding and we show that it successfully reduces the resource requirements of our implementation of Shor's algorithm that is based on the work by~\citet{beauregard03}. Further optimizations could be applied in our IR, for example to quantum circuit definitions or to loop bodies that are free of control flow.

To the best of our knowledge, optimizations that explicitly target mixed quantum-classical programs do not exist yet, in part due to the lack of IRs that are capable of representing such programs. Our IR may thus enable such quantum-classical optimizations. For example, assertion-based optimization may be extended to take branching and loop conditions into account~\cite{haner2020optimization}.

\section{Conclusion}
Our proposed multi-level IR for quantum computing is specifically targeted at quantum-classical co-optimization. In contrast to previous work, it supports carrying out such optimizations at application scale, allowing for optimized resource estimates of large-scale quantum programs. Moreover, QIRO supports quantum-specific optimization passes that may fully leverage the infrastructure provided by MLIR. Crucially, the employed value-semantics in the optimization dialect directly exposes quantum data dependencies. In addition to reuse of existing components, this may enable future quantum program optimizations that leverage dataflow analysis.



\clearpage


\appendix

\section{Mapping Q\# to QIRO}
\label{app:mapping}
In this section, we discuss in detail how to map a quantum program in Q\# to our input dialect.
We choose Q\# as an example front-end for its completeness, most notably with respect to its support for mixed quantum-classical programs.

\subsection{Organization}
Q\# code lives inside (non-nestable) namespaces. We can map these to MLIR modules as they serve a similar purpose.
Furthermore, Q\# allows to bring symbols defined in other namespaces into the current one with an $open$ directive. In this case, the modules should be nested inside the main execution module. When resolving symbols, the front-end should append the corresponding MLIR module identifiers to all references to symbols in those external namespaces, according to the following syntax: \code{@ModuleName::@SymbolName}. The inside of namespaces is composed of global variable definitions, callable definitions such as $operations$ and $functions$, and invocations of callables. How these constructs map to QIRO is described below.

\subsection{Data Types}
Immutable let bindings and mutable variable assignments in Q\# are treated no different in the IR, both which can be mapped to value definition statements. Note that statically, due to MLIR's SSA structure, every value is already defined precisely once.

\subsubsection{Numeric}
MLIR provides standard integer and floating point types of arbitrary bit-width. Literals can be passed to operations that accept arguments in the form of \emph{attributes}, if not, they must first be bound to a value with the \code{constant} op.
Arithmetic expressions can be represented with the appropriate operations from the standard dialect.

\subsubsection{Boolean}
Booleans should be represented by the \code{i1} type.

\subsubsection{Qubit}
Q\#'s Qubit type directly maps to the one present in QIRO. New qubits are created in Q\# with \code{Qubit()}, which translates to a \code{\%q = q.alloc : !q.qubit} operation in our IR.
Once qubit values go out of scope in Q\#, they are automatically deallocated. The front-end should insert explicit \code{q.free \%q : !q.qubit} operations at this point to make the qubit resource available again.

\subsubsection{Arrays}
Note that there is distinction in how Qubit arrays and other arrays are handled.
In general, Q\# can build immutable arrays out of any valid type. These are null initialized, and support slicing and concatenation, which always creates a new array with element copies.
Such classical arrays might be represented by the \code{memref} type, see the MLIR documentation for more details.

Qubit arrays created with \code{Qubit[n]} must instead be represented with the Qureg type from the quantum dialect. The allocation looks as follows: \code{\%r = q.allocreg(n) : !q.qureg<n>}, where the size attribute \code{n} can also be replaced with a dynamic value, in which case the type will not contain a size.
Multidimensional qubit arrays are not supported, so they must be unrolled into a 1D array.
Static bounds checking is implemented where possible, but for dynamic out-of-bounds accesses a run-time exception is expected.
Indexing and slicing is supported at the point of use for quantum registers, as all operations accepting the Qureg type optionally also accept a register access expression for each such argument. It is composed of up to 3 dynamic or static values, corresponding to \emph{start}, \emph{stop}, \emph{step}. If only \emph{start} is present, a single qubit is accessed. If additionally \emph{stop} is present, the slice [\emph{start}, \emph{stop}) with a step of 1 is accessed. An example operation would look like this: \code{q.X \%r[\%a, \%b, 2] : !q.qureg<n>}.
As with qubits, registers need to be explicitly freed when they go out of scope with \code{q.freereg \%r : !q.qureg<n>}.

\subsubsection{Tuples}
While a tuple type is available in MLIR, there are no operations in the standard or quantum dialects that take advantage of them. Functions and circuits have no need for tuples as they are capable of accepting and returning multiple values. Thus the argument and return tuples of Q\# callables should be deconstructed into their components.

\subsubsection{Pauli}
The Pauli type in Q\# is used to indicate rotation axes and measurement bases. For rotations, the corresponding rotation operation should be used \{$Rx$, $Ry$, $Rz$\}. Single qubit measurements in bases other than the Z-basis can be simulated by conjugating the measurement with the corresponding unitary. Joint multi-qubit measurements (e.g. ZX, ZZ, ...) are currently not supported but could easily be added as additional native operations in our IR.

\subsection{Quantum Gates}
Built-in quantum gates (intrinsic operations) in Q\# for the most part have a direct analog in QIRO. If that is not the case, the front-end must express such gates in terms of the provided ones, for which standard algorithms exist.
Measurement is done with a \code{q.measure \%q : !q.qubit} operation in the computational (Z) basis.

\subsubsection{Functors}
Functors in Q\# are the analog to QIRO's meta-operations. These are ``functions'' which take in an operation and produce a new one, modifying its behaviour in the process. The default ones, \emph{adjoint} for inverting an operation and \emph{control} for conditioning the execution based the quantum state of control qubits, are both supported by the quantum dialect.
To use them, the desired operation to be modified must be constructed without target qubits, which produces a value representing the operation. Meta-operations then accept the operation value and the target qubits as arguments. See below for an example of applying an inverted and a controlled rotation gate:
\begin{verbatim}
%R = q.R(%pi) -> !q.u1
q.adj %R, %q : !q.u1, !q.qubit
q.ctrl %R, %c, %t : !q.u1, !q.qubit, !q.qubit
\end{verbatim}

\subsection{Callables}
We deal with two types of callables in Q\#: \emph{functions} and \emph{operations}. Functions contain purely classical code, and intuitively map to functions in MLIR:
\begin{verbatim}
    func @name(arg: argT ..) -> resT.. { ... }
    call @name(arg: argT ..) : (argT..) -> resT..
\end{verbatim}
In contrast, operations contain quantum code (alongside classical one), and this division is exactly reflected by functions and circuits in QIRO. Circuits are similar to functions except that they also operate on quantum data arguments. They can be defined and called as follows:
\begin{verbatim}
    q.circ @name(arg: argT ..) -> resT.. { ... }
    q.call @name(arg: argT ..) : argT.. -> resT..
\end{verbatim}
In order to support meta-operations on circuits, there is also an indirect call mechanism via the \code{apply} operation that operates on the \code{!q.circ} type, or the \code{!q.cop<n, baseT>} type with \code{!q.circ} as its base type.
\begin{verbatim}
    %op = q.getval @name -> !q.circ
    %inv_op = q.adj %op : !q.circ -> !q.circ
    q.apply %inv_op(arg..) : !q.circ(argT..)
\end{verbatim}

\subsection{Conditionals}
Control flow in SSA-based IRs is explicit due to their block structure, which have a single entry and exit point. If/else constructs can easily be represented in MLIR using this block structure and conditional branching.
\begin{verbatim}
cond_br %cond, ^bb1(arg..), ^bb2(arg..) # if
^bb1(arg..):                            # then
    ...
    br ^bb3(arg..)
^bb2(args..):                           # else
    ...
    br ^bb3(arg..)
\end{verbatim}
Here we've shown a simple if-else structure, where \code{\%cond} is a previously calculated boolean condition, and \code{arg..} represents arbitrary block arguments. Arbitrary many else-if sections can be added by inserting more blocks creating conditional branching chains, each with a then block (True) and and a successor block (False) in the chain.

\subsection{Loops}
Index-based for loops in Q\# are well represented by the structured control flow dialect in MLIR. The syntax goes as follows:
\begin{center}
\code{scf.for \%i = <low> to <up> step <step> \{ ... \}},
\end{center}
where the lower/upper bounds and step operands are SSA values.
For-each loops in Q\# can be transformed to this form as well by using the length of the array being traversed as an upper bound. This is one reason that qubit register arguments to circuits always need an accompanying size argument if their size is not statically specified in the type.

Q\#'s while loops (for classical loop conditions) and repeat-until-success loops (for measurement based conditions) are represented by MLIR's standard control flow via blocks and branches. A \code{while (i < n) \{ ...; i++; \}} loop would be translated as follows:
\begin{verbatim}
^bb1(%i: i32):
    ...
    %ip1 = addi %i, %1 : i32
    %cond = cmpi "slt", %ip1, %n : i32
    cond_br %cond, ^bb1(%ip1), ^bb2
\end{verbatim}
where \code{\%1} is the result of the \code{constant} op with the value 1, and \code{\textasciicircum bb2} is the next block after the loop. The entering condition check is omitted for brevity.


\onecolumn
\section{Shor's algorithm in QIRO}
\label{app:shors}
\begin{lstlisting}[language=llvm]
func @mod(%a: i64, %N: i64) -> i64 {
    %0 = divi_unsigned %a, %N : i64
    %1 = muli %N, %0 : i64
    %2 = subi %a, %1 : i64
    return %2 : i64
}

func @mod_exp(%b: i64, %e: i64, %N: i64) -> i64 {
    %c0 = constant 0 : i64
    %c1 = constant 1 : i64
    %c2 = constant 2 : i64
    %cond = cmpi "eq", %N, %c1 : i64
    cond_br %cond, ^ret(%c0 : i64), ^reduce

    ^reduce:
        %res = constant 1 : i64
        %base = call @mod(%b, %N) : (i64, i64) -> i64
        %cond2 = cmpi "ugt", %e, %c0 : i64
        cond_br %cond2, ^while(%base, %e, %res : i64, i64, i64), ^ret(%res : i64)

    ^while(%base_0: i64, %exp_0: i64, %res_0: i64):
        %0 = call @mod(%exp_0, %c2) : (i64, i64) -> i64
        %cond3 = cmpi "eq", %0, %c1 : i64
        %res_1 = scf.if %cond3 -> i64 {
            %1 = muli %res_0, %base_0 : i64
            %2 = call @mod(%1, %N) : (i64, i64) -> i64
            scf.yield %2 : i64
        } else {
            scf.yield %res_0 : i64
        }

        %exp_1 = shift_right_unsigned %exp_0, %c1 : i64

        %3 = muli %base_0, %base_0 : i64
        %base_1 = call @mod(%3, %N) : (i64, i64) -> i64

        %cond4 = cmpi "ugt", %exp_1, %c0 : i64
        cond_br %cond4, ^while(%base_1, %exp_1, %res_1 : i64, i64, i64), ^ret(%res_1 : i64)

    ^ret(%r: i64):
        return %r : i64
}

func @mod_inv(%C: i64, %N: i64) -> i64 {
    %c0 = constant 0 : i64
    %c1 = constant 1 : i64
    br ^while(%N, %C, %c0, %c1 : i64, i64, i64, i64)

    ^while(%r_0: i64, %old_r: i64, %s_0: i64, %old_s: i64):
        %q = divi_unsigned %old_r, %r_0 : i64
        %qr = muli %q, %r_0 : i64
        %r_1 = subi %old_r, %qr : i64

        %qs = muli %q, %s_0 : i64
        %s_1 = subi %old_s, %qs : i64

        %cond = cmpi "ne", %r_1, %c0 : i64
        cond_br %cond, ^while(%r_1, %r_0, %s_1, %s_0 : i64, i64, i64, i64), ^ret(%s_0 : i64)

    ^ret(%s: i64):
        %0 = addi %s, %N : i64
        %1 = call @mod(%0, %N) : (i64, i64) -> i64
        return %1 : i64
}

func @calc_qft_angle(%j: index) -> f64 {
    %pi = constant 3.141592653589793238 : f64
    %c1 = constant 1 : index
    %0 = addi %c1, %j : index
    %1 = shift_left %c1, %0 : index
    %2 = index_cast %1 : index to i64
    %3 = uitofp %2 : i64 to f64
    %4 = divf %pi, %3 : f64
    return %4 : f64
}

func @calc_add_angle(%i: index, %j: index) -> f64 {
    %pi = constant 3.141592653589793238 : f64
    %c1 = constant 1 : index
    %0 = subi %i, %j : index
    %1 = shift_left %c1, %0 : index
    %2 = index_cast %1 : index to i64
    %3 = uitofp %2 : i64 to f64
    %4 = divf %pi, %3 : f64
    return %4 : f64
}

func @calc_cur_a(%N: i64, %n: index, %a: i64, %i: index) -> i64 {
    %c1 = constant 1 : i64
    %c2 = constant 2 : i64
    %k = index_cast %i : index to i64
    %nbits = index_cast %n : index to i64

    %0 = muli %nbits, %c2 : i64
    %1 = subi %0, %c1 : i64
    %2 = subi %1, %k : i64
    %3 = shift_left %c1, %2 : i64
    %4 = call @mod_exp(%a, %3, %N) : (i64, i64, i64) -> i64

    return %4 : i64
}

func @calc_shor_angle(%i: index, %j: index) -> f64 {
    %mpi = constant -3.141592653589793238 : f64
    %c1 = constant 1 : index
    %0 = subi %i, %j : index
    %1 = shift_left %c1, %0 : index
    %2 = index_cast %1 : index to i64
    %3 = uitofp %2 : i64 to f64
    %4 = divf %mpi, %3 : f64
    return %4 : f64
}

// quantum fourier transform on register r
q.circ @QFT(%r: !q.qureg<>, %n : index) attributes {no_inline} {
    %c0 = constant 0 : index
    %c1 = constant 1 : index
    %c2 = constant 2 : index

    scf.for %i = %c0 to %n step %c1 {
        %0 = addi %i, %c1 : index
        %k = subi %n, %0 : index
        q.H %r[%k] : !q.qureg<>
        scf.for %j = %c0 to %k step %c1 {
            %phi = call @calc_qft_angle(%j) : (index) -> f64
            %R = q.R(%phi: f64) -> !q.u1
            %1 = addi %j, %c1 : index
            %h = subi %k, %1 : index
            q.ctrl %R, %r[%h], %r[%k] : !q.u1, !q.qureg<>, !q.qureg<>
        }
    }

    %nd2 = divi_unsigned %n, %c2 : index
    scf.for %i = %c0 to %nd2 step %c1 {
        %0 = addi %i, %c1 : index
        %j = subi %n, %0 : index
        q.SWAP %r[%i], %r[%j] : !q.qureg<>, !q.qureg<>
    }
}

// add a positive or negative number to register of size n
q.circ @addConstant(%C: i64, %r: !q.qureg<>, %n: index) {
    %c0 = constant 0 : index
    %s1 = constant 1 : index
    %c1 = constant 1 : i64

    // compute
    q.call @QFT(%r, %n) {compute} : !q.qureg<>, index

    scf.for %i = %c0 to %n step %s1 {
        %ip1 = addi %i, %s1 : index
        scf.for %j = %c0 to %ip1 step %s1 {
            %k = subi %i, %j : index
            %0 = index_cast %k : index to i64
            %1 = shift_right_signed %C, %0 : i64
            %2 = and %1, %c1 : i64
            %cond = cmpi "eq", %2, %c1 : i64
            scf.if %cond {
                %phi = call @calc_add_angle(%i, %k) : (index, index) -> f64
                q.R(%phi: f64) %r[%i] : !q.qureg<>
            }
        }
    }

    // uncompute
    %qft = q.getval @QFT -> !q.circ
    %qft_inv = q.adj %qft : !q.circ -> !q.circ
    q.apply %qft_inv(%r, %n) {uncompute} : !q.circ(!q.qureg<>, index)
}

// substract a number from register of size n
q.circ @subConstant(%C: i64, %r: !q.qureg<>, %n: index) {
    %cm1 = constant -1 : i64
    %mC = muli %C, %cm1 : i64
    q.call @addConstant(%mC, %r, %n) : i64, !q.qureg<>, index
}

// add a positive number to register modulo N
q.circ @addCmodN(%C: i64, %N: i64, %r: !q.qureg<>, %n: index) {
    %c1 = constant 1 : index
    %nm1 = subi %n, %c1 : index

    q.call @addConstant(%C, %r, %n) : i64, !q.qureg<>, index

    // compute
    q.call @subConstant(%N, %r, %n) {compute} : i64, !q.qureg<>, index
    %anc = q.alloc -> !q.qubit
    q.CX %r[%nm1], %anc {compute} : !q.qureg<>, !q.qubit
    %addOp = q.getval @addConstant -> !q.circ
    %ctrlAdd = q.ctrl %addOp, %anc : !q.circ, !q.qubit -> !q.cop<1, !q.circ>
    q.apply %ctrlAdd(%N, %r, %n) {compute} : !q.cop<1, !q.circ>(i64, !q.qureg<>, index)

    q.call @subConstant(%C, %r, %n) : i64, !q.qureg<>, index

    // uncompute
    q.X %r[%nm1] {uncompute} : !q.qureg<>
    q.CX %r[%nm1], %anc {uncompute} : !q.qureg<>, !q.qubit
    q.X %r[%nm1] {uncompute} : !q.qureg<>
    q.free %anc : !q.qubit

    q.call @addConstant(%C, %r, %n) : i64, !q.qureg<>, index
}

// subtract a positive number to register modulo N
q.circ @subCmodN(%C: i64, %N: i64, %r: !q.qureg<>, %n: index) {
    %NmC = subi %N, %C : i64
    q.call @addCmodN(%NmC, %N, %r, %n) : i64, i64, !q.qureg<>, index
}

// multiply a positive number by a register modulo N, need gcd(C, N) = 1
q.circ @mulCmodN(%C: i64, %N: i64, %r: !q.qureg<>, %n: index) {
    %c0 = constant 0 : index
    %c1 = constant 1 : index
    %np1 = addi %n, %c1 : index
    %anc = q.allocreg(%np1) -> !q.qureg<>
    %Cinv = call @mod_inv(%C, %N) : (i64, i64) -> i64

    scf.for %i = %c0 to %n step %c1 {
        %addOp = q.getval @addCmodN -> !q.circ
        %ctrlAdd = q.ctrl %addOp, %r[%i] : !q.circ, !q.qureg<> -> !q.cop<1, !q.circ>

        %0 = index_cast %i : index to i64
        %1 = shift_left %C, %0 : i64
        %2 = call @mod(%1, %N) : (i64, i64) -> i64
        q.apply %ctrlAdd(%2, %N, %anc, %np1) : !q.cop<1, !q.circ>(i64, i64, !q.qureg<>, index)
    }

    scf.for %i = %c0 to %n step %c1 {
        q.SWAP %anc[%i], %r[%i] : !q.qureg<> , !q.qureg<>
    }

    scf.for %i = %c0 to %n step %c1 {
        %subOp = q.getval @subCmodN -> !q.circ
        %ctrlSub = q.ctrl %subOp, %r[%i] : !q.circ, !q.qureg<> -> !q.cop<1, !q.circ>

        %3 = index_cast %i : index to i64
        %4 = shift_left %Cinv, %3 : i64
        %5 = call @mod(%4, %N) : (i64, i64) -> i64
        q.apply %ctrlSub(%5, %N, %anc, %np1) : !q.cop<1, !q.circ>(i64, i64, !q.qureg<>, index)
    }

    q.freereg %anc : !q.qureg<>
}

q.circ @shor(%N: i64, %a: i64) {
    %c0 = constant 0 : index
    %c1 = constant 1 : index
    %c2 = constant 2 : index

    %0 = uitofp %N : i64 to f64
    %1 = log2 %0 : f64
    %2 = ceilf %1 : f64
    %3 = fptoui %2 : f64 to i64
    %n = index_cast %3 : i64 to index
    %n2 = muli %n, %c2 : index

    %m0 = constant 0 : i1
    %meas = alloc(%n2) : memref<?xi1>
    scf.for %i = %c0 to %n2 step %c1 {
        store %m0, %meas[%i] : memref<?xi1>
    }

    %r = q.allocreg(%n) -> !q.qureg<>
    %cqb = q.alloc -> !q.qubit

    q.X %r[0] : !q.qureg<>

    scf.for %i = %c0 to %n2 step %c1 {
        %cur_a = call @calc_cur_a(%N, %n, %a, %i) : (i64, index, i64, index) -> i64

        q.H %cqb : !q.qubit
        %mulOp = q.getval @mulCmodN -> !q.circ
        %ctrlMul = q.ctrl %mulOp, %cqb : !q.circ, !q.qubit -> !q.cop<1, !q.circ>
        q.apply %ctrlMul(%cur_a, %N, %r, %n) : !q.cop<1, !q.circ>(i64, i64, !q.qureg<>, index)

        scf.for %j = %c0 to %i step %c1 {
            %cond = load %meas[%j] : memref<?xi1>
            scf.if %cond {
                %phi = call @calc_shor_angle(%i, %j) : (index, index) -> f64
                q.R(%phi: f64) %cqb : !q.qubit
            }
        }
        q.H %cqb : !q.qubit

        %m = q.meas %cqb : !q.qubit -> i1
        store %m, %meas[%i] : memref<?xi1>
        scf.if %m {
            q.X %cqb : !q.qubit
        }
    }
    
    %mres = q.meas %r : !q.qureg<> -> memref<?xi1>

    q.free %cqb : !q.qubit
    q.freereg %r : !q.qureg<>

    // process result
}

q.circ @mlir_main(%N : i64, %a : i64) attributes {no_inline_target} {
    q.call @shor(%N, %a) : i64, i64
}
\end{lstlisting}


\begin{thebibliography}{33}


\ifx \showCODEN    \undefined \def \showCODEN     #1{\unskip}     \fi
\ifx \showDOI      \undefined \def \showDOI       #1{#1}\fi
\ifx \showISBNx    \undefined \def \showISBNx     #1{\unskip}     \fi
\ifx \showISBNxiii \undefined \def \showISBNxiii  #1{\unskip}     \fi
\ifx \showISSN     \undefined \def \showISSN      #1{\unskip}     \fi
\ifx \showLCCN     \undefined \def \showLCCN      #1{\unskip}     \fi
\ifx \shownote     \undefined \def \shownote      #1{#1}          \fi
\ifx \showarticletitle \undefined \def \showarticletitle #1{#1}   \fi
\ifx \showURL      \undefined \def \showURL       {\relax}        \fi
\providecommand\bibfield[2]{#2}
\providecommand\bibinfo[2]{#2}
\providecommand\natexlab[1]{#1}
\providecommand\showeprint[2][]{arXiv:#2}

\bibitem[\protect\citeauthoryear{Aho, Lam, Sethi, and Ullman}{Aho
  et~al\mbox{.}}{2006}]%
        {aho86}
\bibfield{author}{\bibinfo{person}{Alfred~V. Aho}, \bibinfo{person}{Monica~S.
  Lam}, \bibinfo{person}{Ravi Sethi}, {and} \bibinfo{person}{Jeffrey~D.
  Ullman}.} \bibinfo{year}{2006}\natexlab{}.
\newblock \bibinfo{booktitle}{\emph{Compilers: Principles, Techniques, and
  Tools (2nd Edition)}}.
\newblock \bibinfo{publisher}{Addison-Wesley}, \bibinfo{address}{USA}.
\newblock


\bibitem[\protect\citeauthoryear{{Amy}, {Maslov}, and {Mosca}}{{Amy}
  et~al\mbox{.}}{2014}]%
        {amy14}
\bibfield{author}{\bibinfo{person}{M. {Amy}}, \bibinfo{person}{D. {Maslov}},
  {and} \bibinfo{person}{M. {Mosca}}.} \bibinfo{year}{2014}\natexlab{}.
\newblock \showarticletitle{Polynomial-Time T-Depth Optimization of Clifford+T
  Circuits Via Matroid Partitioning}.
\newblock \bibinfo{journal}{\emph{IEEE Transactions on Computer-Aided Design of
  Integrated Circuits and Systems}} \bibinfo{volume}{33}, \bibinfo{number}{10}
  (\bibinfo{year}{2014}), \bibinfo{pages}{1476--1489}.
\newblock
\urldef\tempurl%
\url{https://doi.org/10.1109/TCAD.2014.2341953}
\showDOI{\tempurl}


\bibitem[\protect\citeauthoryear{Amy, Maslov, Mosca, and Roetteler}{Amy
  et~al\mbox{.}}{2013}]%
        {amy13}
\bibfield{author}{\bibinfo{person}{Matthew Amy}, \bibinfo{person}{Dmitri
  Maslov}, \bibinfo{person}{Michele Mosca}, {and} \bibinfo{person}{Martin
  Roetteler}.} \bibinfo{year}{2013}\natexlab{}.
\newblock \showarticletitle{A Meet-in-the-Middle Algorithm for Fast Synthesis
  of Depth-Optimal Quantum Circuits}.
\newblock \bibinfo{journal}{\emph{IEEE Transactions on Computer-Aided Design of
  Integrated Circuits and Systems}} \bibinfo{volume}{32}, \bibinfo{number}{6}
  (\bibinfo{date}{Jun} \bibinfo{year}{2013}), \bibinfo{pages}{818--830}.
\newblock
\showISSN{1937-4151}
\urldef\tempurl%
\url{https://doi.org/10.1109/tcad.2013.2244643}
\showDOI{\tempurl}


\bibitem[\protect\citeauthoryear{Beauregard}{Beauregard}{2003}]%
        {beauregard03}
\bibfield{author}{\bibinfo{person}{Stephane Beauregard}.}
  \bibinfo{year}{2003}\natexlab{}.
\newblock \showarticletitle{Circuit for Shor's Algorithm Using 2n+3 Qubits}.
\newblock \bibinfo{journal}{\emph{Quantum Info. Comput.}} \bibinfo{volume}{3},
  \bibinfo{number}{2} (\bibinfo{date}{March} \bibinfo{year}{2003}),
  \bibinfo{pages}{175–185}.
\newblock
\showISSN{1533-7146}
\urldef\tempurl%
\url{https://dl.acm.org/doi/10.5555/2011517.2011525}
\showURL{%
\tempurl}


\bibitem[\protect\citeauthoryear{Bichsel, Baader, Gehr, and Vechev}{Bichsel
  et~al\mbox{.}}{2020}]%
        {silq2020}
\bibfield{author}{\bibinfo{person}{Benjamin Bichsel},
  \bibinfo{person}{Maximilian Baader}, \bibinfo{person}{Timon Gehr}, {and}
  \bibinfo{person}{Martin Vechev}.} \bibinfo{year}{2020}\natexlab{}.
\newblock \showarticletitle{Silq: A High-Level Quantum Language with Safe
  Uncomputation and Intuitive Semantics}. In
  \bibinfo{booktitle}{\emph{Proceedings of the 41st ACM SIGPLAN Conference on
  Programming Language Design and Implementation}} (London, UK)
  \emph{(\bibinfo{series}{PLDI 2020})}. \bibinfo{publisher}{Association for
  Computing Machinery}, \bibinfo{address}{New York, NY, USA},
  \bibinfo{pages}{286–300}.
\newblock
\showISBNx{9781450376136}
\urldef\tempurl%
\url{https://doi.org/10.1145/3385412.3386007}
\showDOI{\tempurl}


\bibitem[\protect\citeauthoryear{Developers}{Developers}{2020}]%
        {cirq2020}
\bibfield{author}{\bibinfo{person}{Cirq Developers}.}
  \bibinfo{year}{2020}\natexlab{}.
\newblock \bibinfo{booktitle}{\emph{Cirq}}.
\newblock
\urldef\tempurl%
\url{https://doi.org/10.5281/zenodo.4062499}
\showDOI{\tempurl}


\bibitem[\protect\citeauthoryear{Draper}{Draper}{2000}]%
        {draper2000}
\bibfield{author}{\bibinfo{person}{Thomas~G Draper}.}
  \bibinfo{year}{2000}\natexlab{}.
\newblock \bibinfo{title}{Addition on a quantum computer}.
\newblock
\newblock
\showeprint[arxiv]{quant-ph/0008033}


\bibitem[\protect\citeauthoryear{Forest, Gosset, Kliuchnikov, and
  McKinnon}{Forest et~al\mbox{.}}{2015}]%
        {forest15}
\bibfield{author}{\bibinfo{person}{Simon Forest}, \bibinfo{person}{David
  Gosset}, \bibinfo{person}{Vadym Kliuchnikov}, {and} \bibinfo{person}{David
  McKinnon}.} \bibinfo{year}{2015}\natexlab{}.
\newblock \showarticletitle{Exact synthesis of single-qubit unitaries over
  Clifford-cyclotomic gate sets}.
\newblock \bibinfo{journal}{\emph{J. Math. Phys.}} \bibinfo{volume}{56},
  \bibinfo{number}{8} (\bibinfo{year}{2015}), \bibinfo{pages}{082201}.
\newblock
\urldef\tempurl%
\url{https://doi.org/10.1063/1.4927100}
\showDOI{\tempurl}


\bibitem[\protect\citeauthoryear{Gambetta, Rodríguez, de~la Puente~González,
  Treinish, Javadi-Abhari, Kassebaum, Pistoia, Hu, tigerjack, Azaustre, Minev,
  Scholten, Oud, Dartiailh, Tod, Cruz-Benito, Wood, and Frisch}{Gambetta
  et~al\mbox{.}}{2019}]%
        {qiskit19}
\bibfield{author}{\bibinfo{person}{Jay Gambetta}, \bibinfo{person}{Diego~M.
  Rodríguez}, \bibinfo{person}{Salvador de~la Puente~González},
  \bibinfo{person}{Matthew Treinish}, \bibinfo{person}{Ali Javadi-Abhari},
  \bibinfo{person}{Paul Kassebaum}, \bibinfo{person}{Marco Pistoia},
  \bibinfo{person}{Shaohan Hu}, \bibinfo{person}{tigerjack},
  \bibinfo{person}{Carlos Azaustre}, \bibinfo{person}{Zlatko Minev},
  \bibinfo{person}{Travis~L. Scholten}, \bibinfo{person}{Steven Oud},
  \bibinfo{person}{Matthieu Dartiailh}, \bibinfo{person}{Maddy Tod},
  \bibinfo{person}{Juan Cruz-Benito}, \bibinfo{person}{Christopher~J. Wood},
  {and} \bibinfo{person}{Albert Frisch}.} \bibinfo{year}{2019}\natexlab{}.
\newblock \bibinfo{booktitle}{\emph{Qiskit: An Open-source Framework for
  Quantum Computing}}.
\newblock
\urldef\tempurl%
\url{https://doi.org/10.5281/zenodo.2573505}
\showDOI{\tempurl}


\bibitem[\protect\citeauthoryear{Geller}{Geller}{2020}]%
        {qir2020}
\bibfield{author}{\bibinfo{person}{Alan Geller}.}
  \bibinfo{year}{2020}\natexlab{}.
\newblock \bibinfo{booktitle}{\emph{Introducing Quantum Intermediate
  Representation (QIR)}}.
\newblock Microsoft.
\newblock
\urldef\tempurl%
\url{https://devblogs.microsoft.com/qsharp/introducing-quantum-intermediate-representation-qir}
\showURL{%
\tempurl}


\bibitem[\protect\citeauthoryear{Gidney and Eker{\aa}}{Gidney and
  Eker{\aa}}{2019}]%
        {gidney19}
\bibfield{author}{\bibinfo{person}{Craig Gidney} {and} \bibinfo{person}{Martin
  Eker{\aa}}.} \bibinfo{year}{2019}\natexlab{}.
\newblock \bibinfo{title}{How to factor 2048 bit rsa integers in 8 hours using
  20 million noisy qubits}.
\newblock
\newblock
\showeprint{1905.09749}~[quant-ph]


\bibitem[\protect\citeauthoryear{Green, Lumsdaine, Ross, Selinger, and
  Valiron}{Green et~al\mbox{.}}{2013}]%
        {quipper13}
\bibfield{author}{\bibinfo{person}{Alexander~S. Green},
  \bibinfo{person}{Peter~LeFanu Lumsdaine}, \bibinfo{person}{Neil~J. Ross},
  \bibinfo{person}{Peter Selinger}, {and} \bibinfo{person}{Beno\^{\i}t
  Valiron}.} \bibinfo{year}{2013}\natexlab{}.
\newblock \showarticletitle{Quipper: A Scalable Quantum Programming Language}.
  In \bibinfo{booktitle}{\emph{Proceedings of the 34th ACM SIGPLAN Conference
  on Programming Language Design and Implementation}} (Seattle, Washington,
  USA) \emph{(\bibinfo{series}{PLDI '13})}. \bibinfo{publisher}{Association for
  Computing Machinery}, \bibinfo{address}{New York, NY, USA},
  \bibinfo{pages}{333–342}.
\newblock
\showISBNx{9781450320146}
\urldef\tempurl%
\url{https://doi.org/10.1145/2491956.2462177}
\showDOI{\tempurl}


\bibitem[\protect\citeauthoryear{H\"{a}ner, Hoefler, and Troyer}{H\"{a}ner
  et~al\mbox{.}}{2020}]%
        {haner2020optimization}
\bibfield{author}{\bibinfo{person}{Thomas H\"{a}ner}, \bibinfo{person}{Torsten
  Hoefler}, {and} \bibinfo{person}{Matthias Troyer}.}
  \bibinfo{year}{2020}\natexlab{}.
\newblock \showarticletitle{Assertion-Based Optimization of Quantum Programs}.
\newblock \bibinfo{journal}{\emph{Proc. ACM Program. Lang.}}
  \bibinfo{volume}{4}, \bibinfo{number}{OOPSLA}, Article
  \bibinfo{articleno}{133} (\bibinfo{date}{Nov.} \bibinfo{year}{2020}),
  \bibinfo{numpages}{20}~pages.
\newblock
\urldef\tempurl%
\url{https://doi.org/10.1145/3428201}
\showDOI{\tempurl}


\bibitem[\protect\citeauthoryear{H{\"a}ner, Jaques, Naehrig, Roetteler, and
  Soeken}{H{\"a}ner et~al\mbox{.}}{2020}]%
        {haner2020elliptic}
\bibfield{author}{\bibinfo{person}{Thomas H{\"a}ner}, \bibinfo{person}{Samuel
  Jaques}, \bibinfo{person}{Michael Naehrig}, \bibinfo{person}{Martin
  Roetteler}, {and} \bibinfo{person}{Mathias Soeken}.}
  \bibinfo{year}{2020}\natexlab{}.
\newblock \showarticletitle{Improved Quantum Circuits for Elliptic Curve
  Discrete Logarithms}. In \bibinfo{booktitle}{\emph{Post-Quantum
  Cryptography}}, \bibfield{editor}{\bibinfo{person}{Jintai Ding} {and}
  \bibinfo{person}{Jean-Pierre Tillich}} (Eds.). \bibinfo{publisher}{Springer
  International Publishing}, \bibinfo{address}{Cham},
  \bibinfo{pages}{425--444}.
\newblock
\showISBNx{978-3-030-44223-1}
\urldef\tempurl%
\url{https://doi.org/10.1007/978-3-030-44223-1_23}
\showDOI{\tempurl}


\bibitem[\protect\citeauthoryear{H{\"a}ner, Steiger, Svore, and
  Troyer}{H{\"a}ner et~al\mbox{.}}{2018}]%
        {haner18}
\bibfield{author}{\bibinfo{person}{Thomas H{\"a}ner}, \bibinfo{person}{Damian~S
  Steiger}, \bibinfo{person}{Krysta Svore}, {and} \bibinfo{person}{Matthias
  Troyer}.} \bibinfo{year}{2018}\natexlab{}.
\newblock \showarticletitle{A software methodology for compiling quantum
  programs}.
\newblock \bibinfo{journal}{\emph{Quantum Science and Technology}}
  \bibinfo{volume}{3}, \bibinfo{number}{2} (\bibinfo{year}{2018}),
  \bibinfo{pages}{020501}.
\newblock
\urldef\tempurl%
\url{https://doi.org/10.1088/2058-9565/aaa5cc}
\showDOI{\tempurl}


\bibitem[\protect\citeauthoryear{Iten, Colbeck, Kukuljan, Home, and
  Christandl}{Iten et~al\mbox{.}}{2016}]%
        {iten16}
\bibfield{author}{\bibinfo{person}{Raban Iten}, \bibinfo{person}{Roger
  Colbeck}, \bibinfo{person}{Ivan Kukuljan}, \bibinfo{person}{Jonathan Home},
  {and} \bibinfo{person}{Matthias Christandl}.}
  \bibinfo{year}{2016}\natexlab{}.
\newblock \showarticletitle{Quantum circuits for isometries}.
\newblock \bibinfo{journal}{\emph{Phys. Rev. A}}  \bibinfo{volume}{93}
  (\bibinfo{date}{Mar} \bibinfo{year}{2016}), \bibinfo{pages}{032318}.
\newblock
Issue 3.
\urldef\tempurl%
\url{https://doi.org/10.1103/PhysRevA.93.032318}
\showDOI{\tempurl}


\bibitem[\protect\citeauthoryear{JavadiAbhari, Faruque, Dousti, Svec, Catu,
  Chakrabati, Chiang, Vanderwilt, Black, Chong, Martonosi, Suchara, Brown,
  Pedram, and Brun}{JavadiAbhari et~al\mbox{.}}{2012}]%
        {scaffold12}
\bibfield{author}{\bibinfo{person}{Ali JavadiAbhari}, \bibinfo{person}{Arvin
  Faruque}, \bibinfo{person}{Mohammad~Javad Dousti}, \bibinfo{person}{Lukas
  Svec}, \bibinfo{person}{Oana Catu}, \bibinfo{person}{Amlan Chakrabati},
  \bibinfo{person}{Chen-Fu Chiang}, \bibinfo{person}{Seth Vanderwilt},
  \bibinfo{person}{John Black}, \bibinfo{person}{Fred Chong},
  \bibinfo{person}{Margaret Martonosi}, \bibinfo{person}{Martin Suchara},
  \bibinfo{person}{Ken Brown}, \bibinfo{person}{Massoud Pedram}, {and}
  \bibinfo{person}{Todd Brun}.} \bibinfo{year}{2012}\natexlab{}.
\newblock \bibinfo{booktitle}{\emph{Scaffold: Quantum programming language}}.
\newblock \bibinfo{type}{{T}echnical {R}eport}. \bibinfo{institution}{Princeton
  University}.
\newblock
\urldef\tempurl%
\url{https://www.cs.princeton.edu/research/techreps/TR-934-12}
\showURL{%
\tempurl}


\bibitem[\protect\citeauthoryear{JavadiAbhari, Patil, Kudrow, Heckey, Lvov,
  Chong, and Martonosi}{JavadiAbhari et~al\mbox{.}}{2014}]%
        {scaffcc14}
\bibfield{author}{\bibinfo{person}{Ali JavadiAbhari}, \bibinfo{person}{Shruti
  Patil}, \bibinfo{person}{Daniel Kudrow}, \bibinfo{person}{Jeff Heckey},
  \bibinfo{person}{Alexey Lvov}, \bibinfo{person}{Frederic~T. Chong}, {and}
  \bibinfo{person}{Margaret Martonosi}.} \bibinfo{year}{2014}\natexlab{}.
\newblock \showarticletitle{ScaffCC: A Framework for Compilation and Analysis
  of Quantum Computing Programs}. In \bibinfo{booktitle}{\emph{Proceedings of
  the 11th ACM Conference on Computing Frontiers}} (Cagliari, Italy)
  \emph{(\bibinfo{series}{CF '14})}. \bibinfo{publisher}{Association for
  Computing Machinery}, \bibinfo{address}{New York, NY, USA}, Article
  \bibinfo{articleno}{1}, \bibinfo{numpages}{10}~pages.
\newblock
\showISBNx{9781450328708}
\urldef\tempurl%
\url{https://doi.org/10.1145/2597917.2597939}
\showDOI{\tempurl}


\bibitem[\protect\citeauthoryear{Killoran, Izaac, Quesada, Bergholm, Amy, and
  Weedbrook}{Killoran et~al\mbox{.}}{2019}]%
        {strawberryfields19}
\bibfield{author}{\bibinfo{person}{Nathan Killoran}, \bibinfo{person}{Josh
  Izaac}, \bibinfo{person}{Nicol{\'{a}}s Quesada}, \bibinfo{person}{Ville
  Bergholm}, \bibinfo{person}{Matthew Amy}, {and} \bibinfo{person}{Christian
  Weedbrook}.} \bibinfo{year}{2019}\natexlab{}.
\newblock \showarticletitle{Strawberry {F}ields: {A} {S}oftware {P}latform for
  {P}hotonic {Q}uantum {C}omputing}.
\newblock \bibinfo{journal}{\emph{{Quantum}}}  \bibinfo{volume}{3}
  (\bibinfo{date}{March} \bibinfo{year}{2019}), \bibinfo{pages}{129}.
\newblock
\showISSN{2521-327X}
\urldef\tempurl%
\url{https://doi.org/10.22331/q-2019-03-11-129}
\showDOI{\tempurl}


\bibitem[\protect\citeauthoryear{{Lattner} and {Adve}}{{Lattner} and
  {Adve}}{2004}]%
        {llvm04}
\bibfield{author}{\bibinfo{person}{C. {Lattner}} {and} \bibinfo{person}{V.
  {Adve}}.} \bibinfo{year}{2004}\natexlab{}.
\newblock \showarticletitle{LLVM: a compilation framework for lifelong program
  analysis \& transformation}. In \bibinfo{booktitle}{\emph{International
  Symposium on Code Generation and Optimization, 2004. CGO 2004.}}
  \bibinfo{publisher}{IEEE}, \bibinfo{address}{San Jose, CA, USA},
  \bibinfo{pages}{75--86}.
\newblock
\urldef\tempurl%
\url{https://doi.org/10.1109/CGO.2004.1281665}
\showDOI{\tempurl}


\bibitem[\protect\citeauthoryear{{Lattner}, {Amini}, {Bondhugula}, {Cohen},
  {Davis}, {Pienaar}, {Riddle}, {Shpeisman}, {Vasilache}, and
  {Zinenko}}{{Lattner} et~al\mbox{.}}{2021}]%
        {mlir2021}
\bibfield{author}{\bibinfo{person}{C. {Lattner}}, \bibinfo{person}{M. {Amini}},
  \bibinfo{person}{U. {Bondhugula}}, \bibinfo{person}{A. {Cohen}},
  \bibinfo{person}{A. {Davis}}, \bibinfo{person}{J. {Pienaar}},
  \bibinfo{person}{R. {Riddle}}, \bibinfo{person}{T. {Shpeisman}},
  \bibinfo{person}{N. {Vasilache}}, {and} \bibinfo{person}{O. {Zinenko}}.}
  \bibinfo{year}{2021}\natexlab{}.
\newblock \showarticletitle{MLIR: Scaling Compiler Infrastructure for Domain
  Specific Computation}. In \bibinfo{booktitle}{\emph{2021 IEEE/ACM
  International Symposium on Code Generation and Optimization (CGO)}}.
  \bibinfo{publisher}{IEEE}, \bibinfo{address}{Seoul, Korea (South)},
  \bibinfo{pages}{2--14}.
\newblock
\urldef\tempurl%
\url{https://doi.org/10.1109/CGO51591.2021.9370308}
\showDOI{\tempurl}


\bibitem[\protect\citeauthoryear{McCaskey and Nguyen}{McCaskey and
  Nguyen}{2021}]%
        {mccaskey2021}
\bibfield{author}{\bibinfo{person}{Alexander McCaskey} {and}
  \bibinfo{person}{Thien Nguyen}.} \bibinfo{year}{2021}\natexlab{}.
\newblock \bibinfo{title}{A MLIR Dialect for Quantum Assembly Languages}.
\newblock
\newblock
\showeprint[arxiv]{2101.11365}~[quant-ph]


\bibitem[\protect\citeauthoryear{Meuli, Soeken, Roetteler, and H\"{a}ner}{Meuli
  et~al\mbox{.}}{2020}]%
        {meuli2020}
\bibfield{author}{\bibinfo{person}{Giulia Meuli}, \bibinfo{person}{Mathias
  Soeken}, \bibinfo{person}{Martin Roetteler}, {and} \bibinfo{person}{Thomas
  H\"{a}ner}.} \bibinfo{year}{2020}\natexlab{}.
\newblock \showarticletitle{Enabling Accuracy-Aware Quantum Compilers Using
  Symbolic Resource Estimation}.
\newblock \bibinfo{journal}{\emph{Proc. ACM Program. Lang.}}
  \bibinfo{volume}{4}, \bibinfo{number}{OOPSLA}, Article
  \bibinfo{articleno}{130} (\bibinfo{date}{Nov.} \bibinfo{year}{2020}),
  \bibinfo{numpages}{26}~pages.
\newblock
\urldef\tempurl%
\url{https://doi.org/10.1145/3428198}
\showDOI{\tempurl}


\bibitem[\protect\citeauthoryear{Nam, Ross, Su, Childs, and Maslov}{Nam
  et~al\mbox{.}}{2018}]%
        {nam18}
\bibfield{author}{\bibinfo{person}{Yunseong Nam}, \bibinfo{person}{Neil~J
  Ross}, \bibinfo{person}{Yuan Su}, \bibinfo{person}{Andrew~M Childs}, {and}
  \bibinfo{person}{Dmitri Maslov}.} \bibinfo{year}{2018}\natexlab{}.
\newblock \showarticletitle{Automated optimization of large quantum circuits
  with continuous parameters}.
\newblock \bibinfo{journal}{\emph{npj Quantum Information}}
  \bibinfo{volume}{4}, \bibinfo{number}{1} (\bibinfo{year}{2018}),
  \bibinfo{pages}{1--12}.
\newblock
\urldef\tempurl%
\url{https://doi.org/10.1038/s41534-018-0072-4}
\showDOI{\tempurl}


\bibitem[\protect\citeauthoryear{Nielsen and Chuang}{Nielsen and
  Chuang}{2010}]%
        {nielsen02}
\bibfield{author}{\bibinfo{person}{Michael~A. Nielsen} {and}
  \bibinfo{person}{Isaac~L. Chuang}.} \bibinfo{year}{2010}\natexlab{}.
\newblock \bibinfo{booktitle}{\emph{Quantum Computation and Quantum
  Information: 10th Anniversary Edition}}.
\newblock \bibinfo{publisher}{Cambridge University Press},
  \bibinfo{address}{UK}.
\newblock
\urldef\tempurl%
\url{https://doi.org/10.1017/CBO9780511976667}
\showDOI{\tempurl}


\bibitem[\protect\citeauthoryear{Paetznick and Svore}{Paetznick and
  Svore}{2014}]%
        {paetznick13}
\bibfield{author}{\bibinfo{person}{Adam Paetznick} {and}
  \bibinfo{person}{Krysta~M. Svore}.} \bibinfo{year}{2014}\natexlab{}.
\newblock \showarticletitle{Repeat-until-Success: Non-Deterministic
  Decomposition of Single-Qubit Unitaries}.
\newblock \bibinfo{journal}{\emph{Quantum Info. Comput.}} \bibinfo{volume}{14},
  \bibinfo{number}{15–16} (\bibinfo{date}{Nov.} \bibinfo{year}{2014}),
  \bibinfo{pages}{1277–1301}.
\newblock
\showISSN{1533-7146}
\urldef\tempurl%
\url{https://dl.acm.org/doi/10.5555/2685179.2685181}
\showURL{%
\tempurl}


\bibitem[\protect\citeauthoryear{Paykin, Rand, and Zdancewic}{Paykin
  et~al\mbox{.}}{2017}]%
        {qwire17}
\bibfield{author}{\bibinfo{person}{Jennifer Paykin}, \bibinfo{person}{Robert
  Rand}, {and} \bibinfo{person}{Steve Zdancewic}.}
  \bibinfo{year}{2017}\natexlab{}.
\newblock \showarticletitle{QWIRE: A Core Language for Quantum Circuits}. In
  \bibinfo{booktitle}{\emph{Proceedings of the 44th ACM SIGPLAN Symposium on
  Principles of Programming Languages}} (Paris, France)
  \emph{(\bibinfo{series}{POPL 2017})}. \bibinfo{publisher}{Association for
  Computing Machinery}, \bibinfo{address}{New York, NY, USA},
  \bibinfo{pages}{846–858}.
\newblock
\showISBNx{9781450346603}
\urldef\tempurl%
\url{https://doi.org/10.1145/3009837.3009894}
\showDOI{\tempurl}


\bibitem[\protect\citeauthoryear{Preskill}{Preskill}{2018}]%
        {preskill18}
\bibfield{author}{\bibinfo{person}{John Preskill}.}
  \bibinfo{year}{2018}\natexlab{}.
\newblock \showarticletitle{Quantum {C}omputing in the {NISQ} era and beyond}.
\newblock \bibinfo{journal}{\emph{{Quantum}}}  \bibinfo{volume}{2}
  (\bibinfo{date}{Aug.} \bibinfo{year}{2018}), \bibinfo{pages}{79}.
\newblock
\showISSN{2521-327X}
\urldef\tempurl%
\url{https://doi.org/10.22331/q-2018-08-06-79}
\showDOI{\tempurl}


\bibitem[\protect\citeauthoryear{Reiher, Wiebe, Svore, Wecker, and
  Troyer}{Reiher et~al\mbox{.}}{2017}]%
        {reiher16}
\bibfield{author}{\bibinfo{person}{Markus Reiher}, \bibinfo{person}{Nathan
  Wiebe}, \bibinfo{person}{Krysta~M. Svore}, \bibinfo{person}{Dave Wecker},
  {and} \bibinfo{person}{Matthias Troyer}.} \bibinfo{year}{2017}\natexlab{}.
\newblock \showarticletitle{Elucidating reaction mechanisms on quantum
  computers}.
\newblock \bibinfo{journal}{\emph{Proceedings of the National Academy of
  Sciences}} \bibinfo{volume}{114}, \bibinfo{number}{29}
  (\bibinfo{year}{2017}), \bibinfo{pages}{7555--7560}.
\newblock
\showISSN{0027-8424}
\urldef\tempurl%
\url{https://doi.org/10.1073/pnas.1619152114}
\showDOI{\tempurl}


\bibitem[\protect\citeauthoryear{Smith, Curtis, and Zeng}{Smith
  et~al\mbox{.}}{2016}]%
        {quil16}
\bibfield{author}{\bibinfo{person}{Robert~S. Smith},
  \bibinfo{person}{Michael~J. Curtis}, {and} \bibinfo{person}{William~J.
  Zeng}.} \bibinfo{year}{2016}\natexlab{}.
\newblock \bibinfo{title}{A Practical Quantum Instruction Set Architecture}.
\newblock
\newblock
\showeprint{1608.03355}~[quant-ph]


\bibitem[\protect\citeauthoryear{Steiger, H{\"a}ner, and Troyer}{Steiger
  et~al\mbox{.}}{2018}]%
        {projectq18}
\bibfield{author}{\bibinfo{person}{Damian~S. Steiger}, \bibinfo{person}{Thomas
  H{\"a}ner}, {and} \bibinfo{person}{Matthias Troyer}.}
  \bibinfo{year}{2018}\natexlab{}.
\newblock \showarticletitle{ProjectQ: an open source software framework for
  quantum computing}.
\newblock \bibinfo{journal}{\emph{Quantum}}  \bibinfo{volume}{2}
  (\bibinfo{date}{Jan} \bibinfo{year}{2018}), \bibinfo{pages}{49}.
\newblock
\showISSN{2521-327X}
\urldef\tempurl%
\url{https://doi.org/10.22331/q-2018-01-31-49}
\showDOI{\tempurl}


\bibitem[\protect\citeauthoryear{Svore, Geller, Troyer, Azariah, Granade, Heim,
  Kliuchnikov, Mykhailova, Paz, and Roetteler}{Svore et~al\mbox{.}}{2018}]%
        {qsharp18}
\bibfield{author}{\bibinfo{person}{Krysta Svore}, \bibinfo{person}{Alan
  Geller}, \bibinfo{person}{Matthias Troyer}, \bibinfo{person}{John Azariah},
  \bibinfo{person}{Christopher Granade}, \bibinfo{person}{Bettina Heim},
  \bibinfo{person}{Vadym Kliuchnikov}, \bibinfo{person}{Mariia Mykhailova},
  \bibinfo{person}{Andres Paz}, {and} \bibinfo{person}{Martin Roetteler}.}
  \bibinfo{year}{2018}\natexlab{}.
\newblock \showarticletitle{Q\#: Enabling Scalable Quantum Computing and
  Development with a High-Level DSL}. In \bibinfo{booktitle}{\emph{Proceedings
  of the Real World Domain Specific Languages Workshop 2018}} (Vienna, Austria)
  \emph{(\bibinfo{series}{RWDSL2018})}. \bibinfo{publisher}{Association for
  Computing Machinery}, \bibinfo{address}{New York, NY, USA}, Article
  \bibinfo{articleno}{7}, \bibinfo{numpages}{10}~pages.
\newblock
\showISBNx{9781450363556}
\urldef\tempurl%
\url{https://doi.org/10.1145/3183895.3183901}
\showDOI{\tempurl}


\bibitem[\protect\citeauthoryear{von Burg, Low, H{\"a}ner, Steiger, Reiher,
  Roetteler, and Troyer}{von Burg et~al\mbox{.}}{2020}]%
        {vonburg2020}
\bibfield{author}{\bibinfo{person}{Vera von Burg}, \bibinfo{person}{Guang~Hao
  Low}, \bibinfo{person}{Thomas H{\"a}ner}, \bibinfo{person}{Damian~S Steiger},
  \bibinfo{person}{Markus Reiher}, \bibinfo{person}{Martin Roetteler}, {and}
  \bibinfo{person}{Matthias Troyer}.} \bibinfo{year}{2020}\natexlab{}.
\newblock \bibinfo{title}{Quantum computing enhanced computational catalysis}.
\newblock
\newblock
\showeprint{2007.14460}~[quant-ph]


\end{thebibliography}
\end{document}